\documentclass[aps,prl,twocolumn,amsmath,amssymb,superscriptaddress,floatfix]{revtex4-2}
\usepackage{graphicx,bm,mathtools,hyperref,xcolor}

\hypersetup{colorlinks=true,linkcolor=blue,citecolor=blue,urlcolor=blue}

\begin{document}

\title{Certifying Hidden Dissipation from\\
       Observed Current Fluctuations}

\author{Ahmed Roman}
\affiliation{Department of Medical Oncology, Dana-Farber Cancer Institute,
Boston, Massachusetts 02215, USA}
\affiliation{Broad Institute of MIT and Harvard,
Cambridge, Massachusetts 02142, USA}
\affiliation{Harvard Medical School, Boston, Massachusetts 02115, USA}

\date{September 11, 2026}

\begin{abstract}
A single-molecule experiment on a driven enzyme or motor resolves
only a few of the transitions the machine makes, yet one would like
to know how much free energy it dissipates in total, including on the
steps that stay hidden. The mean and the fluctuations of the currents
on the watched transitions are shown to certify this hidden
dissipation, with no knowledge of the transition rates and no access
to the hidden transitions: when the watched transitions span the
cycles of the network, the observed-current covariance recovers the
full density-contracted quadratic current geometry, including the contribution of
transitions that are never seen. The mechanism is that fluctuations
are set by the dynamical activity, or traffic, the symmetric partner
of the current. Contracting the large-deviation cost of empirical
currents over density fluctuations identifies the observed-current
covariance with a traffic-weighted metric, and turns partial
observation into a minimum-energy completion problem over the unseen
cycles, whose solution is the certified hidden cost. Existing
current-fluctuation uncertainty relations bound dissipation from
chosen currents but do not say when partial observation fixes the
hidden contribution; the cycle-observability condition derived here
does. The statements concern long-time means and fluctuations;
finite-time estimates require separate error control.
\end{abstract}

\maketitle
\begin{figure}[t]
\includegraphics[width=\columnwidth]{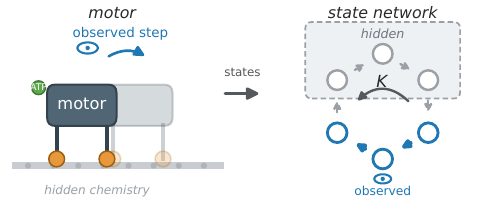}
\caption{From a single-molecule trajectory to a dissipation bound.
A molecular motor steps along its track by consuming fuel: the
mechanical step produces a visible displacement (observed), while the
chemical transitions that drive it are not resolved (hidden). The
same dynamics is a continuous-time Markov chain on discrete states
(right), in which a few transitions are watched (solid blue) and the
rest are hidden (dashed, within the shaded panel), with stationary
cycle current $K$. The mean and the fluctuations of the watched
currents bound the total entropy production $\sigma$ from below by
$\mathcal{B}_{\mathcal{O}}^{\max}$, including the dissipation on
transitions that are never observed.}
\label{fig:schematic}
\end{figure}

\paragraph*{Introduction.}
Consider a molecular motor such as kinesin stepping along a
microtubule, or an enzyme cycling through its conformational states.
Each runs through a network of discrete states, hopping between them
at rates biased by the chemical fuel that drives it, and a
single-molecule experiment typically resolves only a few of these
transitions (the mechanical steps that produce a visible
displacement, say), while the underlying chemical substeps stay
hidden [Fig.~\ref{fig:schematic}]. How much free energy is the whole
machine dissipating, including on the steps one never sees?

This Letter answers that question for finite-state Markov jump systems
in a nonequilibrium steady state. Watching the mean flow and the
fluctuations on a chosen handful of transitions is enough to bound
the total entropy production from below, with no knowledge of the
transition rates and no access to the hidden transitions; the bound
tightens as more transitions are watched and can even certify
dissipation carried entirely by transitions that are never observed.
The essential ingredient beyond the mean currents is their
\emph{fluctuations}, whose size encodes geometric information about
the hidden part of the network.

The reason fluctuations carry this information is that they are
shaped by two distinct features of the dynamics. Persistent
probability currents distinguish nonequilibrium steady
states from equilibrium ones, but currents are only the antisymmetric
part of the stationary flux. The symmetric part, the dynamical
activity or traffic, fixes the kinetic scale: it determines how much
back-and-forth motion accompanies a given net current. Two steady
states with the same stationary density and current can therefore
differ in fluctuations, relaxation, and entropy production because
their traffic differs.

This kinetic sector is visible in current fluctuations.
Thermodynamic uncertainty relations and related inference schemes
\cite{Gingrich2016,Pietzonka2016,PLE2016,DTB2019,Maes2017,
Manikandan2020,BNF2024,DTBM2025} bound entropy production from
passive measurements of empirical currents, using their means and
covariances. In these approaches the covariance enters as an
empirical object. Its geometric origin, and its behavior under
partial edge observation, remain less explicit.

The observed-current covariance therefore has a precise variational
meaning, which is the technical heart of the ``traffic energy'' picture
above. Contracting the level-2.5 large-deviation rate function
\cite{BFG2015,MN2007} over normalized empirical-density fluctuations
leaves a metric $M_T$ on the cycle-current subspace, with inverse
traffic $1/T_e$ as the local edge-current quadratic. The pseudoinverse
of the observed-edge covariance equals the corresponding effective
metric obtained by minimizing $M_T$-energy over hidden cycle-current
completions. Partial observation thus does not merely discard edges; it
induces a minimum-energy completion problem. The central consequence
is a certification theorem. When the watched transitions determine
the cycle currents, the completion has a unique solution, and the
observed-current covariance recovers the full $M_T$-energy of the
network, including the contribution of edges that are never observed;
when the physical stationary current is unscreened,
$\gamma_G(K)=0$, this equals the raw quadratic traffic cost
$\sum_e K_e^2/T_e$, where sums over $e$ count each undirected edge
once. Existing
current-fluctuation relations bound dissipation from a chosen current
but do not identify when partial observation pins down the hidden
part. Combining the covariance with the exact nonlinear Schnakenberg
cost on the observed edges then gives a lower bound on entropy
production that is monotone under adding observed edges.

\paragraph*{Related work.}
Inference under partial information has been approached in several ways.
Shiraishi and Sagawa \cite{ShiraishiSagawa2015} gave a fluctuation
theorem for partially masked dynamics, Polettini and Esposito
\cite{PLE2017} an effective thermodynamics for a marginal observer, and
Bisker \emph{et al.}\ \cite{Bisker2017} hierarchical bounds from partial
information. Edge- and transition-resolved schemes infer dissipation
from the statistics of a few visible transitions \cite{Harunari2022},
from waiting-time distributions \cite{vanderMeer2022,Ertel2024}, and as
a tightest hidden-entropy bound \cite{Ehrich2021}. These use
coarse-grained, marginal, or waiting-time descriptions; the present
bound instead uses observed edge currents and their covariance.

A complementary line bounds dissipation from current fluctuations: the
multivariate large-deviation bound of Gingrich \emph{et al.}\
\cite{Gingrich2016} and its refinements \cite{Pietzonka2016,PLE2016,Dechant2018,VVH2020},
kinetic and frenetic sharpenings \cite{DTB2019,Maes2017}, finite-time
optimization over currents \cite{Manikandan2020}, and recent
partial-data methods \cite{BNF2024,DTBM2025}. The optimized multivariate
relation bounds dissipation from chosen currents but does not say when
observed edges force a unique hidden-current completion. Waiting-time
methods also infer hidden cycles \cite{vanderMeer2022}; the present
result instead gives a
covariance-based cycle-observability condition that certifies hidden
cycle energy from observed edges, combined with the exact observed-edge
Schnakenberg cost into a single monotone bound.

\paragraph*{Setup.}
Let an irreducible continuous-time Markov chain on $N$ states have
transition rates $w_{ij} \ge 0$, stationary distribution $p_i > 0$,
and stationary one-way fluxes $q_{ij} = p_i w_{ij}$. Assume bidirectional
support: $w_{ij}>0$ if and only if $w_{ji}>0$. For each
undirected edge $e = \{i,j\}$ in the edge set $\mathcal{E}$, fix an
orientation $i \to j$ and define
\begin{equation}
K_e = q_{ij} - q_{ji}, \qquad
T_e = q_{ij} + q_{ji}, \qquad
\eta_e = \frac{K_e}{T_e},
\end{equation}
with $|\eta_e| < 1$. $K_e$ is the antisymmetric stationary current,
which together with $p$ characterizes the nonequilibrium steady state
in the density-current classification of Zia and Schmittmann
\cite{ZS2007}; $T_e$ is the symmetric activity (traffic), and
$\eta_e$ the dimensionless local irreversibility. Stationarity is
$B K = 0$, where $B \in
\mathbb{R}^{N \times |\mathcal{E}|}$ is the incidence matrix; a cycle
basis $C \in \mathbb{R}^{|\mathcal{E}| \times r}$ with $r =
|\mathcal{E}| - N + 1$ and $B C = 0$ writes any divergence-free
current as $K = C\alpha$. Given $(p, K, T)$ the rates are
\begin{equation}
w_{ij} = \frac{T_e + K_e}{2 p_i},
\end{equation}
so the triple $(p, K, T)$ fixes a representative generator when the
underlying graph is known. Below, $T$ also supplies the local metric
of the empirical-current covariance. The Schnakenberg edge entropy production rate is
\begin{equation}
\sigma_e \;=\; K_e \log\!\left( \frac{T_e + K_e}{T_e - K_e} \right)
        \;=\; 2 T_e\, \eta_e \,\mathrm{artanh}(\eta_e).
\label{eq:edge-sigma}
\end{equation}
Total dissipation and total stationary jump activity are $\sigma =
\sum_e \sigma_e$ and $\kappa = \sum_e T_e$.

\paragraph*{Question.}
A nonequilibrium steady state can be inferred from passive current
statistics. Gingrich \emph{et al.}~\cite{Gingrich2016} proved a
quadratic upper bound on the level-2.5 rate function for
empirical currents and contracted it onto the scalar empirical current
$\hat j_d = \langle d, \hat j \rangle$. Here $d=(d_e)$ is a fixed
vector of edge weights and $\hat j$ is the vector of net oriented jump
counts per unit time, with stationary mean $j=K$. The resulting
thermodynamic uncertainty relation is $\sigma \ge 2j_d^2/\Sigma_d$,
where $j_d=\langle d,j\rangle$ and
$\Sigma_d=\lim_{t\to\infty}t\,\mathrm{Var}(\hat j_d)$. The bound uses
the mean and variance of the chosen current; transition rates and
external control are not required. The covariance object that appears
on the data side has been used in subsequent inference work
\cite{PLE2016,Pietzonka2016,Bisker2017,DTB2019,Maes2017}. The natural
question concerns the geometric origin of the metric in that
covariance: which features of the underlying chain control its size,
and how does it connect to the symmetric activity sector?

\paragraph*{Quadratic contraction.}
At long time $t$, the joint fluctuations of empirical density
$\hat\rho$ and empirical flow $\hat\phi_{ij}$ satisfy a large-deviation
principle with speed $t$ and rate function $I[\rho,\phi]$, the
level-2.5 functional \cite{BFG2015,MN2007} restricted to stationary
continuity $\sum_j (\phi_{ij} - \phi_{ji}) = 0$. Around the typical
point $\rho = p$, $\phi_{ij} = q_{ij}$, expansion to second order
gives
\begin{equation}
I^{(2)}(\delta\rho, \delta\phi) = \tfrac12 \sum_{i,j}
  \frac{(\delta\phi_{ij} - w_{ij}\, \delta\rho_i)^2}{q_{ij}},
\label{eq:level25}
\end{equation}
with the sum over directed pairs. Per undirected edge, change
variables to $\delta K_e = \delta\phi_{ij} - \delta\phi_{ji}$ and
$\delta T_e = \delta\phi_{ij} + \delta\phi_{ji}$, and define
$(G\, \delta\rho)_e \equiv w_{ij}\, \delta\rho_i - w_{ji}\, \delta\rho_j$
for the chosen orientation. The covariance formulas below are
invariant under orientation reversal because $K_e$, $\delta K_e$, and
$(G\delta\rho)_e$ all change sign together. Eq.~\eqref{eq:level25}
contains $\delta T_e$ unconstrained. Minimizing over $\delta T_e$ at
fixed $\delta K_e$ and $\delta\rho$ collapses each edge term to
$\bigl(\delta K_e - (G\delta\rho)_e\bigr)^2 / (2T_e)$, giving
\begin{equation}
I^{(2)}(\delta K, \delta\rho) = \tfrac12 \bigl(\delta K - G\delta\rho\bigr)^\top
  D_T^{-1} \bigl(\delta K - G\delta\rho\bigr),
\label{eq:I-after-T}
\end{equation}
with $D_T = \mathrm{diag}(T_e)$. Density fluctuations are normalized: $\mathbf{1}^\top \delta\rho = 0$.
Let $R \in \mathbb{R}^{N \times (N-1)}$ be a basis for
$\mathbf{1}^\perp$, so $\delta\rho = R\xi$ with $\xi \in \mathbb{R}^{N-1}$.
Minimizing over $\xi$ leaves a quadratic form on $\delta K$ alone:
\begin{equation}
I^{(2)}_{\rm cyc}(\delta K) = \tfrac12 \, \delta K^\top M_T \, \delta K,
\label{eq:cyc-rate}
\end{equation}
with the density-contracted traffic metric
\begin{equation}
M_T \;=\; D_T^{-1} \;-\; D_T^{-1} G R\,
        (R^\top G^\top D_T^{-1} G R)^+\, R^\top G^\top D_T^{-1}.
\label{eq:MT}
\end{equation}
Below, $\tilde G \equiv G R$ denotes $G$ restricted to normalized
density fluctuations; the screening identities involve $\tilde G$
in place of $G$.
Restricting to divergence-free currents $\delta K = C\, \delta\alpha$
and inverting on the cycle subspace, the asymptotic CLT covariance of
empirical edge currents, projected onto the cycle subspace, is
\begin{equation}
\Sigma_K \;=\; C\, (C^\top M_T\, C)^{-1}\, C^\top.
\label{eq:Sigma}
\end{equation}
Here $\Sigma_K=\lim_{t\to\infty}t\,\mathrm{Cov}(\hat K)$, so
$\sqrt t(\hat K-K)$ has limiting covariance $\Sigma_K$.
$\Sigma_K$ has rank $r$ and vanishes on the divergence-carrying
complement; finite-time empirical edge currents are divergence-free
only up to boundary terms, and \eqref{eq:Sigma} is the long-time
Gaussian limit. Thus $D_T^{-1}$ is the local traffic metric before
density contraction, while $M_T$ is the metric seen by empirical
currents after the empirical density has relaxed optimally.

\paragraph*{Screening interpretation of $G$.}
Equation \eqref{eq:cyc-rate} can be rewritten as
\begin{equation}
\delta K^\top M_T\, \delta K
= \min_{\xi \in \mathbb{R}^{N-1}}
  (\delta K - \tilde G\, \xi)^\top D_T^{-1}\,
  (\delta K - \tilde G\, \xi),
\label{eq:screening}
\end{equation}
so $\tilde G$ identifies the directions in edge-current space along
which an apparent fluctuation can be matched by a shift in normalized
empirical occupations rather than by a genuine cycle current. The
difference
\begin{align}
\Delta_G(K) &=\; K^\top D_T^{-1} K - K^\top M_T K \\
&=\; K^\top D_T^{-1} \tilde G\,
       (\tilde G^\top D_T^{-1} \tilde G)^+\,
       \tilde G^\top D_T^{-1} K \;\ge\; 0
\label{eq:screened-cost}
\end{align}
is the traffic cost screened by density relaxation, in analogy with
electrostatic screening. For the physical current, $\gamma_G(K)=0$
is equivalent to $\tilde G^\top D_T^{-1}K=0$. The stronger condition
$C^\top D_T^{-1} \tilde G = 0$ means that screening vanishes for
every cycle current, equivalently
$M_T |_\mathcal{C} = D_T^{-1} |_\mathcal{C}$. Below, the
cycle-observability limit recovers $K^\top M_T K$ from observed-edge
covariance, which equals the raw cost $K^\top D_T^{-1} K$ when the
physical current is unscreened.

Let $\mathcal{O} \subseteq \mathcal{E}$ denote an observed edge
subset, $P_{\mathcal{O}} : \mathbb{R}^{|\mathcal{E}|} \to
\mathbb{R}^{|\mathcal{O}|}$ the coordinate projection, and
$\Sigma_{\mathcal{O}} = P_{\mathcal{O}} \Sigma_K
P_{\mathcal{O}}^\top$, $K_{\mathcal{O}} = P_{\mathcal{O}} K$ the
asymptotic mean and covariance of the observed currents. Throughout,
unhatted symbols are exact long-time objects; hats denote finite-time
plug-in estimates $\widehat K_{\mathcal{O}}(t)$,
$\widehat\Sigma_{\mathcal{O}}(t)$.

\paragraph*{Covariance metric and the completion problem.}
The Hessian of \eqref{eq:cyc-rate} is $M_T$, so $\Sigma_K = C(C^\top
M_T C)^{-1} C^\top$ as in \eqref{eq:Sigma}. The scalar TUR
\cite{Gingrich2016} bounds dissipation by any one current,
$\sigma \ge 2(u^\top K)^2/(u^\top \Sigma_K u)$; the sharpest such
bound, taken over current directions $u$, is
$\sigma \ge 2 K^\top \Sigma_K^+ K$, since $K^\top \Sigma_K^+ K =
\sup_{u^\top\Sigma_Ku>0}(u^\top K)^2/(u^\top \Sigma_K u)$. On the cycle subspace
this optimized envelope is exactly the density-contracted traffic
energy, $K^\top \Sigma_K^+ K = K^\top M_T K \le K^\top D_T^{-1} K$,
the inequality following from $M_T \preceq D_T^{-1}$. The data-side
covariance thus measures a traffic energy, not an abstract variance.

When only the subset $\mathcal{O}$ is observed, the same energy
becomes a constrained minimum. The Gaussian projection identity,
applied with metric $M_T|_{\mathcal{C}}$ and projection
$P_{\mathcal{O}}$, gives
\begin{equation}
\boxed{\;
K_{\mathcal{O}}^\top \Sigma_{\mathcal{O}}^+ K_{\mathcal{O}}
\;=\; \min_{\substack{J \in \mathcal{C}\\
                       P_{\mathcal{O}} J = K_{\mathcal{O}}}}
      J^\top M_T J.
\;}
\label{eq:min-energy}
\end{equation}
The observed-current covariance quadratic form equals the least
density-contracted traffic energy of any stationary current that
reproduces those currents on $\mathcal{O}$. Partial observation does
not discard the hidden edges; it replaces them by their cheapest
admissible completion.

Two limits make this completion sharp. If the observed edges span
cycle space ($P_{\mathcal{O}}$ injective on $\mathcal{C}$), the only
admissible completion is $K$ itself, so the observed covariance
returns the full network energy $K_{\mathcal{O}}^\top
\Sigma_{\mathcal{O}}^+ K_{\mathcal{O}} = K^\top M_T K$. If in addition
the physical current is unscreened, $\gamma_G(K)=0$, this equals the raw cost
$\sum_{e\in\mathcal{E}} K_e^2/T_e$, which strictly exceeds the
observed-edge sum $\sum_{e\in\mathcal{O}} K_e^2/T_e$ whenever a hidden
edge carries current. Computing the lower bound uses only observed
means and covariances; this exact-recovery certification additionally
requires cycle observability, equivalently
$\operatorname{rank}\Sigma_{\mathcal O}=r$ when $r$ is known
(Supplemental Material).

\paragraph*{Hybrid bound.}
The covariance term is quadratic, while the exact edge cost
\eqref{eq:edge-sigma} is not. On observed edges the full nonlinear
cost is available, since $K_e$ and $T_e$ both follow from the forward
and backward counts. Writing $\sigma_e = 2K_e^2/T_e + R_e$ with
$R_e = K_e\log[(T_e+K_e)/(T_e-K_e)] - 2K_e^2/T_e \ge 0$, the observed
remainders add to the covariance completion without spoiling the
lower bound,
\begin{equation}
\sigma \;\ge\; \mathcal{B}_{\mathcal{O}}
   \;\equiv\; 2\, K_{\mathcal{O}}^\top \Sigma_{\mathcal{O}}^+
              K_{\mathcal{O}}
   \;+\; \sum_{e \in \mathcal{O}} R_e,
\label{eq:hybrid}
\end{equation}
because $\sigma = 2K^\top D_T^{-1}K + \sum_{\mathcal{E}} R_e \ge
2K^\top M_T K + \sum_{\mathcal{O}} R_e \ge \mathcal{B}_{\mathcal{O}}$.
The per-edge nonlinear sum $\sum_{\mathcal{O}} \sigma_e$ is itself a
lower bound, so the operational estimator is the larger of the two,
\begin{equation}
\boxed{\;
\sum_{e \in \mathcal{O}} \sigma_e \;\le\;
\mathcal{B}_{\mathcal{O}}^{\max} \equiv
\max\!\Big\{\textstyle\sum_{e \in \mathcal{O}} \sigma_e,\,
\mathcal{B}_{\mathcal{O}}\Big\} \;\le\; \sigma.
\;}
\label{eq:bound}
\end{equation}
Adding observed edges shrinks the feasible set in
\eqref{eq:min-energy} and adds nonnegative remainders, so
$\mathcal{B}_{\mathcal{O}}^{\max}$ never decreases as more transitions
are watched. Formal statements of these results, with the projection
and screening derivations, are collected in the Supplemental
Material \cite{SuppMat}.

Every quantity needed to evaluate \eqref{eq:bound} comes from the
observed jumps; the hidden network and the transition rates enter
nowhere. On each observed edge the forward and backward jump counts
fix $\hat K_e$, $\hat T_e$, $\hat\sigma_e$, and $\hat R_e$, while the
current covariance is estimated by blocking the trajectory; the
explicit recipe, with the two free parameters (block length and
pseudoinverse cutoff) that control sampling error but not the
inequality, is given in the Supplemental Material.

\paragraph*{What is new and what is inherited.}
The scalar-to-multivariate optimization of the TUR is not new:
maximizing the scalar TUR $\sigma \ge 2 (u^\top K)^2/(u^\top \Sigma_K
u)$ over $u$ yields $\sigma \ge 2 K^\top \Sigma_K^+ K$, which is
implicit in current-fluctuation inference \cite{Gingrich2016,PLE2016}.
The contributions of the present work are:
(i) the cycle-observability condition and the strict hidden-edge
advantage, under which the observed-current covariance certifies the
full hidden quadratic cost, a statement
absent from the optimized TUR, which bounds dissipation from chosen
currents but does not characterize when partial observation fixes the
hidden contribution;
(ii) the partial-edge minimum-completion identity \eqref{eq:min-energy},
which recasts the covariance estimator as the least-energy stationary
completion of hidden currents and supplies the mechanism behind (i);
(iii) the level-2.5 contraction identifying the metric in that
covariance object as the density-contracted traffic metric $M_T$
\eqref{eq:MT}, with $\tilde G^\top D_T^{-1}K=0$ characterizing
unscreened physical currents and $C^\top D_T^{-1}\tilde G=0$
characterizing absence of screening on the entire cycle subspace;
(iv) the nonlinear correction $\sum_{\mathcal{O}} R_e$ tightening the
bound to $\mathcal{B}_{\mathcal{O}}$ and combining with the per-edge
sum into the monotone operational estimator
$\mathcal{B}_{\mathcal{O}}^{\max}$.
The cycle quadratic bound in Appendix~B of Ref.~\cite{GRH2017}
uses force-weighted conductances $F_e/(2 K_e)$ with
$F_e=\log(q_{ij}/q_{ji})=\log(w_{ij}p_i/w_{ji}p_j)$; the expansion
$F_e/(2 K_e)=(1/T_e)[1 + \eta_e^2/3 + O(\eta_e^4)]$ shows the force
conductance is the nonlinear entropy metric and $1/T_e$ its leading
covariance form.

\paragraph*{Demonstration: validity regime and screening obstruction.}
For $K\ne0$, the natural control parameter is the dimensionless screening fraction
\begin{equation}
\gamma_G(K) \;\equiv\; \frac{\Delta_G(K)}{K^\top D_T^{-1} K}, \qquad
\Delta_G(K) = K^\top (D_T^{-1} - M_T) K,
\label{eq:gamma}
\end{equation}
which controls the full-observation hybrid gap
$\sigma - \mathcal{B}_{\mathcal{E}} = 2 \Delta_G(K) =
2 \gamma_G(K) \cdot K^\top D_T^{-1} K$. At $\gamma_G(K) = 0$ and full
observation, the covariance completion equals the raw quadratic traffic
cost and the hybrid bound saturates $\sigma$. At finite $\gamma_G(K)$,
the full-observation hybrid level is $\sigma-2\Delta_G(K)$.
The operational estimator always satisfies
$\mathcal B_{\mathcal E}^{\max}=\sigma$, because it includes the
full observed-edge sum.

\begin{figure*}[tbp]
\includegraphics[width=\textwidth]{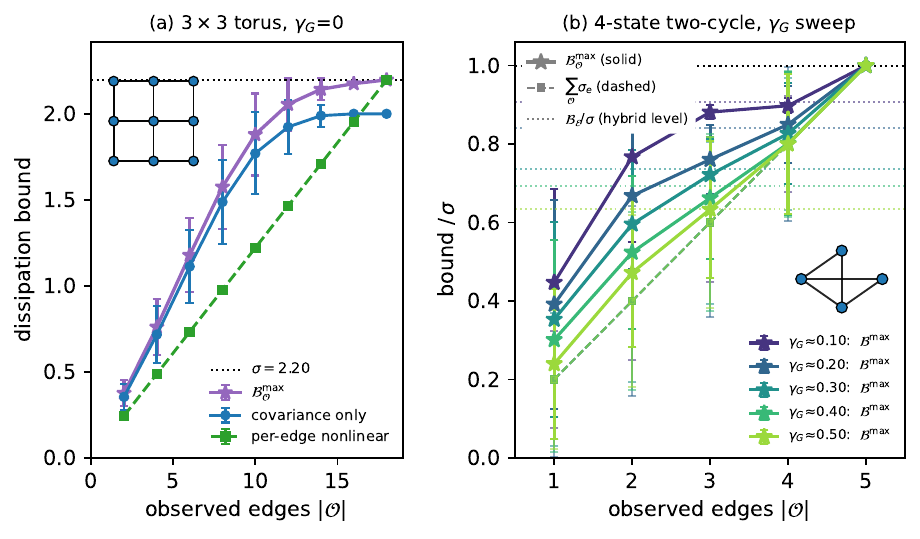}
\caption{Exact-theory dissipation bounds; all quantities are
asymptotic long-time CLT objects.
(a) $3 \times 3$ torus, $\gamma_G(K) = 0$: no screening of the
physical stationary current.
The three estimators (per-edge nonlinear sum,
covariance term, operational
$\mathcal{B}_{\mathcal{O}}^{\max}$) are plotted against the number
of observed edges, averaged over $400$ random subsets per size
(all subsets if at most $500$), with
error bars equal to subset standard deviations. The hybrid curve is
omitted because its mean differs from the operational bound by less
than $0.02\%$ of $\sigma$ at every plotted coverage. The operational
estimator saturates $\sigma$ at full coverage and recovers $93.5\%$
of $\sigma$ when one third of the edges are hidden.
(b) Four-state, five-edge two-cycle network with nonuniform driving,
$\gamma_G(K) \in \{0.10, 0.20, 0.30, 0.40, 0.50\}$, found by numerical
search in rate space. Solid stars trace
$\mathcal{B}_{\mathcal{O}}^{\max}/\sigma$; dashed squares trace
$\sum_{\mathcal{O}}\sigma_e/\sigma$; dotted horizontal lines mark
the predicted full-observation hybrid level $1 - 2\gamma_G(K) (K^\top
D_T^{-1} K)/\sigma$ for each $\gamma_G(K)$. Curves enumerate all
$\binom{5}{n}$ observed subsets at every $n$. The degradation with
$\gamma_G(K)$ is not a failure of the bound but the predicted loss of
raw quadratic traffic cost under density-current screening.}
\label{fig:main}
\end{figure*}

Figure~\ref{fig:main} resolves both regimes from exact
theory.~Panel~(a) takes the no-screening reference, the $3 \times 3$
torus ($N = 9$, $|\mathcal{E}| = 18$, cycle rank ten) with uniform
stationary measure and biased rates giving $\eta_{\max} \approx 0.5$,
$\sigma = 2.20$, and $\gamma_G(K) = 0$. The per-edge sum, covariance
term, and operational bound in \eqref{eq:bound} are computed using the analytic $K$, $T$, and
$\Sigma_K = C(C^\top M_T C)^{-1} C^\top$ over $400$ random observed
subsets per size, or all subsets when their number is at most $500$;
the operational $\mathcal{B}_{\mathcal{O}}^{\max}$ saturates
$\sigma$ at full coverage and degrades only mildly under hiding,
while the per-edge sum drops in proportion to the observed fraction.
Panel~(b) sweeps the four-state two-cycle network with nonuniform
driving over $\gamma_G(K) \in \{0.10, 0.2, 0.30, 0.4, 0.50\}$, found by numerical
search in rate space. For each $\gamma_G(K)$, all $\binom{5}{n}$
observed subsets at every size are enumerated. Plotted normalized to
$\sigma$, the operational $\mathcal{B}_{\mathcal{O}}^{\max}$ tracks
the per-edge sum at high coverage and the hybrid term at low
coverage; at full observation it equals one. The dotted levels
$1 - 2\gamma_G(K) \cdot (K^\top D_T^{-1} K)/\sigma$ instead mark the
full-observation hybrid bounds. Screening reduces the covariance
contribution, while the operational estimator retains the observed-edge
entropy sum.

\paragraph*{Finite-time and statistics.}
The bounds above are stated for exact long-time means $K$ and CLT
covariances $\Sigma_K$. Finite-time plug-in estimators
$\widehat K_{\mathcal{O}}(t)$, $\widehat\Sigma_{\mathcal{O}}(t)$
can be biased by covariance inversion and finite-length blocking;
statistical error control is separate from the thermodynamic inequality.
Fixed blocks retain a finite-window bias even as $t\to\infty$. A
controlled test on a four-state two-cycle network (where two observed
edges (a cotree) certify the full energy of all five, including the
three hidden ones, and long-block estimates approximate this level) is
given in the Supplemental Material.

\begin{figure}[!htbp]
\includegraphics[width=\columnwidth]{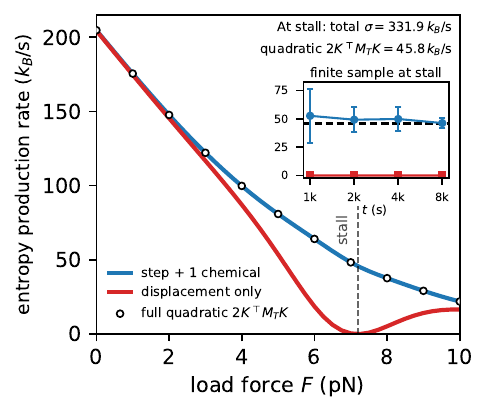}
\caption{Certification on the Liepelt--Lipowsky kinesin network
\cite{LL2007,LL2009} ($1\,$mM ATP, $1\,\mu$M ADP/P$_{\mathrm i}$). Certified
entropy production rate $2 K_{\mathcal{O}}^\top \Sigma_{\mathcal{O}}^+ K_{\mathcal{O}}$
versus load force: displacement-only observation (red) collapses at
the stall force ($\approx 7.2\,$pN), where ATP is still hydrolyzed but
the net step current vanishes, while step plus one chemical transition
(blue) certifies the full quadratic cost $2 K^\top M_T K$ (black open
circles, coincident with the blue curve) at every load. At stall,
the total entropy production rate is $\sigma=331.9\,k_B/\mathrm{s}$
and the full quadratic cost is $45.8\,k_B/\mathrm{s}$; multiplying
these rates by the bath temperature gives the corresponding powers
in $k_BT/\mathrm{s}$. Inset: finite-sample estimate at stall from simulated
trajectories; the step-plus-chemical plug-in (blue, mean $\pm$ s.d.\
over $14$ trajectories) approximates the exact value (black dashed), while
the displacement-only estimate (red) stays near zero. Blocks of
$50\,$s retain a $0.3\%$ population bias.}
\label{fig:kinesin}
\end{figure}

\paragraph*{Application to kinesin.}
The condition has teeth on a real motor. We take the six-state
low-product-concentration chemomechanical network of Liepelt and
Lipowsky \cite{LL2007,LL2009}, whose
literature parametrization combines experimental data with balance
constraints (conventions in the Supplement); one transition is the mechanical step that a displacement
detector resolves, the rest are chemical and hidden. The cycle space
has rank two: a forward and a backward stepping cycle, both through
the mechanical edge, and a futile hydrolysis cycle that consumes ATP
with no net step and contains no mechanical edge. Observing the step
alone is therefore not cycle-observing, and the futile dissipation is
invisible to the displacement-current bound. Figure~\ref{fig:kinesin} traces the
certified energy against load force. As the stall force is approached,
the forward and backward stepping cycles balance, the net step current
vanishes, and the displacement-only certificate collapses to zero even
though the motor keeps hydrolyzing ATP. Adding a single chemical
transition restores cycle observability and certifies the full cycle
energy at every load. The inset illustrates recovery at stall from
simulated trajectories: the step-plus-chemical estimate approaches the
quadratic level within sampling error, while the displacement-only
estimate stays near zero.

\paragraph*{Discussion.}
Two facts combine. The Schnakenberg relation \eqref{eq:edge-sigma}
makes $T_e$ the kinetic denominator of edge dissipation, with $1/T_e$
the leading covariance form of the force-weighted entropy metric of
current-fluctuation inference \cite{Gingrich2016,PLE2016}. The
level-2.5 contraction \eqref{eq:cyc-rate} turns $D_T^{-1}$, after
eliminating density fluctuations, into the cycle metric $M_T$, and the
completion identity \eqref{eq:min-energy} makes
$K_{\mathcal{O}}^\top \Sigma_{\mathcal{O}}^+ K_{\mathcal{O}}$ the least
density-contracted traffic energy consistent with the observed
currents. This equals the full network energy $K^\top M_T K$ when the
observed edges span cycle space. The completion is a Schur complement
of the cycle metric; when $C^\top D_T^{-1}\tilde G=0$ it is Thomson's principle with edge
resistances $1/T_e$, an effective-resistance reduction of the hidden
subnetwork.

What the experiment must supply is the long-time current covariance,
not the rates: the bound is computable and valid without hidden
topology, which enters only to certify exact recovery. Covariance
estimation requires many blocks long compared with current-correlation
times, not merely the fast substeps that dominate the traffic.
The screening fraction
$\gamma_G(K)$ measures how much raw traffic cost the density sector
absorbs; it vanishes by symmetry for rate-transitive networks, so the
gap between the certified and the raw cost is itself a structural
property of the chain.

\begin{acknowledgments}
\interlinepenalty=10000
The author is a Damon Runyon Quantitative Biology Fellow.
OpenAI's Codex was used for assistance with code checks, error-bar cleanup,
and editorial revisions.
\end{acknowledgments}

\paragraph*{Data availability.} Code and numerical data are available at
\url{https://github.com/AhmedHRoman/Roman-Numerals}.




\bibliographystyle{apsrev4-2}
\bibliography{References}

\begin{thebibliography}{24}%
\makeatletter
\providecommand \@ifxundefined [1]{%
 \@ifx{#1\undefined}
}%
\providecommand \@ifnum [1]{%
 \ifnum #1\expandafter \@firstoftwo
 \else \expandafter \@secondoftwo
 \fi
}%
\providecommand \@ifx [1]{%
 \ifx #1\expandafter \@firstoftwo
 \else \expandafter \@secondoftwo
 \fi
}%
\providecommand \natexlab [1]{#1}%
\providecommand \enquote  [1]{``#1''}%
\providecommand \bibnamefont  [1]{#1}%
\providecommand \bibfnamefont [1]{#1}%
\providecommand \citenamefont [1]{#1}%
\providecommand \href@noop [0]{\@secondoftwo}%
\providecommand \href [0]{\begingroup \@sanitize@url \@href}%
\providecommand \@href[1]{\@@startlink{#1}\@@href}%
\providecommand \@@href[1]{\endgroup#1\@@endlink}%
\providecommand \@sanitize@url [0]{\catcode `\\12\catcode `\$12\catcode
  `\&12\catcode `\#12\catcode `\^12\catcode `\_12\catcode `\%12\relax}%
\providecommand \@@startlink[1]{}%
\providecommand \@@endlink[0]{}%
\providecommand \url  [0]{\begingroup\@sanitize@url \@url }%
\providecommand \@url [1]{\endgroup\@href {#1}{\urlprefix }}%
\providecommand \urlprefix  [0]{URL }%
\providecommand \Eprint [0]{\href }%
\providecommand \doibase [0]{https://doi.org/}%
\providecommand \selectlanguage [0]{\@gobble}%
\providecommand \bibinfo  [0]{\@secondoftwo}%
\providecommand \bibfield  [0]{\@secondoftwo}%
\providecommand \translation [1]{[#1]}%
\providecommand \BibitemOpen [0]{}%
\providecommand \bibitemStop [0]{}%
\providecommand \bibitemNoStop [0]{.\EOS\space}%
\providecommand \EOS [0]{\spacefactor3000\relax}%
\providecommand \BibitemShut  [1]{\csname bibitem#1\endcsname}%
\let\auto@bib@innerbib\@empty
\bibitem [{\citenamefont {Gingrich}\ \emph {et~al.}(2016)\citenamefont
  {Gingrich}, \citenamefont {Horowitz}, \citenamefont {Perunov},\ and\
  \citenamefont {England}}]{Gingrich2016}%
  \BibitemOpen
  \bibfield  {author} {\bibinfo {author} {\bibfnamefont {T.~R.}\ \bibnamefont
  {Gingrich}}, \bibinfo {author} {\bibfnamefont {J.~M.}\ \bibnamefont
  {Horowitz}}, \bibinfo {author} {\bibfnamefont {N.}~\bibnamefont {Perunov}},\
  and\ \bibinfo {author} {\bibfnamefont {J.~L.}\ \bibnamefont {England}},\
  }\href {https://doi.org/10.1103/PhysRevLett.116.120601} {\bibfield  {journal}
  {\bibinfo  {journal} {Phys. Rev. Lett.}\ }\textbf {\bibinfo {volume} {116}},\
  \bibinfo {pages} {120601} (\bibinfo {year} {2016})}\BibitemShut {NoStop}%
\bibitem [{\citenamefont {Pietzonka}\ \emph {et~al.}(2016)\citenamefont
  {Pietzonka}, \citenamefont {Barato},\ and\ \citenamefont
  {Seifert}}]{Pietzonka2016}%
  \BibitemOpen
  \bibfield  {author} {\bibinfo {author} {\bibfnamefont {P.}~\bibnamefont
  {Pietzonka}}, \bibinfo {author} {\bibfnamefont {A.~C.}\ \bibnamefont
  {Barato}},\ and\ \bibinfo {author} {\bibfnamefont {U.}~\bibnamefont
  {Seifert}},\ }\href {https://doi.org/10.1103/PhysRevE.93.052145} {\bibfield
  {journal} {\bibinfo  {journal} {Phys. Rev. E}\ }\textbf {\bibinfo {volume}
  {93}},\ \bibinfo {pages} {052145} (\bibinfo {year} {2016})}\BibitemShut
  {NoStop}%
\bibitem [{\citenamefont {Polettini}\ \emph {et~al.}(2016)\citenamefont
  {Polettini}, \citenamefont {Lazarescu},\ and\ \citenamefont
  {Esposito}}]{PLE2016}%
  \BibitemOpen
  \bibfield  {author} {\bibinfo {author} {\bibfnamefont {M.}~\bibnamefont
  {Polettini}}, \bibinfo {author} {\bibfnamefont {A.}~\bibnamefont
  {Lazarescu}},\ and\ \bibinfo {author} {\bibfnamefont {M.}~\bibnamefont
  {Esposito}},\ }\href {https://doi.org/10.1103/PhysRevE.94.052104} {\bibfield
  {journal} {\bibinfo  {journal} {Phys. Rev. E}\ }\textbf {\bibinfo {volume}
  {94}},\ \bibinfo {pages} {052104} (\bibinfo {year} {2016})}\BibitemShut
  {NoStop}%
\bibitem [{\citenamefont {Di~Terlizzi}\ and\ \citenamefont
  {Baiesi}(2019)}]{DTB2019}%
  \BibitemOpen
  \bibfield  {author} {\bibinfo {author} {\bibfnamefont {I.}~\bibnamefont
  {Di~Terlizzi}}\ and\ \bibinfo {author} {\bibfnamefont {M.}~\bibnamefont
  {Baiesi}},\ }\href@noop {} {\bibfield  {journal} {\bibinfo  {journal}
  {Journal of Physics A: Mathematical and Theoretical}\ }\textbf {\bibinfo
  {volume} {52}},\ \bibinfo {pages} {02LT03} (\bibinfo {year}
  {2019})}\BibitemShut {NoStop}%
\bibitem [{\citenamefont {Maes}(2017)}]{Maes2017}%
  \BibitemOpen
  \bibfield  {author} {\bibinfo {author} {\bibfnamefont {C.}~\bibnamefont
  {Maes}},\ }\href {https://doi.org/10.1103/PhysRevLett.119.160601} {\bibfield
  {journal} {\bibinfo  {journal} {Phys. Rev. Lett.}\ }\textbf {\bibinfo
  {volume} {119}},\ \bibinfo {pages} {160601} (\bibinfo {year}
  {2017})}\BibitemShut {NoStop}%
\bibitem [{\citenamefont {Manikandan}\ \emph {et~al.}(2020)\citenamefont
  {Manikandan}, \citenamefont {Gupta},\ and\ \citenamefont
  {Krishnamurthy}}]{Manikandan2020}%
  \BibitemOpen
  \bibfield  {author} {\bibinfo {author} {\bibfnamefont {S.~K.}\ \bibnamefont
  {Manikandan}}, \bibinfo {author} {\bibfnamefont {D.}~\bibnamefont {Gupta}},\
  and\ \bibinfo {author} {\bibfnamefont {S.}~\bibnamefont {Krishnamurthy}},\
  }\href {https://doi.org/10.1103/PhysRevLett.124.120603} {\bibfield  {journal}
  {\bibinfo  {journal} {Phys. Rev. Lett.}\ }\textbf {\bibinfo {volume} {124}},\
  \bibinfo {pages} {120603} (\bibinfo {year} {2020})}\BibitemShut {NoStop}%
\bibitem [{\citenamefont {Baiesi}\ \emph {et~al.}(2024)\citenamefont {Baiesi},
  \citenamefont {Nishiyama},\ and\ \citenamefont {Falasco}}]{BNF2024}%
  \BibitemOpen
  \bibfield  {author} {\bibinfo {author} {\bibfnamefont {M.}~\bibnamefont
  {Baiesi}}, \bibinfo {author} {\bibfnamefont {T.}~\bibnamefont {Nishiyama}},\
  and\ \bibinfo {author} {\bibfnamefont {G.}~\bibnamefont {Falasco}},\
  }\href@noop {} {\bibfield  {journal} {\bibinfo  {journal} {Communications
  Physics}\ }\textbf {\bibinfo {volume} {7}},\ \bibinfo {pages} {264} (\bibinfo
  {year} {2024})}\BibitemShut {NoStop}%
\bibitem [{\citenamefont {Di~Terlizzi}(2025)}]{DTBM2025}%
  \BibitemOpen
  \bibfield  {author} {\bibinfo {author} {\bibfnamefont {I.}~\bibnamefont
  {Di~Terlizzi}},\ }\href {https://doi.org/10.1103/fsph-437v} {\bibfield
  {journal} {\bibinfo  {journal} {Phys. Rev. Lett.}\ }\textbf {\bibinfo
  {volume} {135}},\ \bibinfo {pages} {237101} (\bibinfo {year}
  {2025})}\BibitemShut {NoStop}%
\bibitem [{\citenamefont {Bertini}\ \emph {et~al.}(2015)\citenamefont
  {Bertini}, \citenamefont {Faggionato},\ and\ \citenamefont
  {Gabrielli}}]{BFG2015}%
  \BibitemOpen
  \bibfield  {author} {\bibinfo {author} {\bibfnamefont {L.}~\bibnamefont
  {Bertini}}, \bibinfo {author} {\bibfnamefont {A.}~\bibnamefont
  {Faggionato}},\ and\ \bibinfo {author} {\bibfnamefont {D.}~\bibnamefont
  {Gabrielli}},\ }in\ \href@noop {} {\emph {\bibinfo {booktitle} {Annales de
  l'IHP Probabilit{\'e}s et statistiques}}},\ Vol.~\bibinfo {volume} {51}\
  (\bibinfo {year} {2015})\ pp.\ \bibinfo {pages} {867--900}\BibitemShut
  {NoStop}%
\bibitem [{\citenamefont {Maes}\ and\ \citenamefont
  {Netočný}(2007)}]{MN2007}%
  \BibitemOpen
  \bibfield  {author} {\bibinfo {author} {\bibfnamefont {C.}~\bibnamefont
  {Maes}}\ and\ \bibinfo {author} {\bibfnamefont {K.}~\bibnamefont
  {Netočný}},\ }\href {https://doi.org/10.1016/j.crhy.2007.05.003} {\bibfield
   {journal} {\bibinfo  {journal} {Comptes Rendus Physique}\ }\textbf {\bibinfo
  {volume} {8}},\ \bibinfo {pages} {591} (\bibinfo {year} {2007})},\ \bibinfo
  {note} {work, dissipation, and fluctuations in nonequilibrium
  physics}\BibitemShut {NoStop}%
\bibitem [{\citenamefont {Shiraishi}\ and\ \citenamefont
  {Sagawa}(2015)}]{ShiraishiSagawa2015}%
  \BibitemOpen
  \bibfield  {author} {\bibinfo {author} {\bibfnamefont {N.}~\bibnamefont
  {Shiraishi}}\ and\ \bibinfo {author} {\bibfnamefont {T.}~\bibnamefont
  {Sagawa}},\ }\href {https://doi.org/10.1103/PhysRevE.91.012130} {\bibfield
  {journal} {\bibinfo  {journal} {Phys. Rev. E}\ }\textbf {\bibinfo {volume}
  {91}},\ \bibinfo {pages} {012130} (\bibinfo {year} {2015})}\BibitemShut
  {NoStop}%
\bibitem [{\citenamefont {Polettini}\ and\ \citenamefont
  {Esposito}(2017)}]{PLE2017}%
  \BibitemOpen
  \bibfield  {author} {\bibinfo {author} {\bibfnamefont {M.}~\bibnamefont
  {Polettini}}\ and\ \bibinfo {author} {\bibfnamefont {M.}~\bibnamefont
  {Esposito}},\ }\href {https://doi.org/10.1103/PhysRevLett.119.240601}
  {\bibfield  {journal} {\bibinfo  {journal} {Phys. Rev. Lett.}\ }\textbf
  {\bibinfo {volume} {119}},\ \bibinfo {pages} {240601} (\bibinfo {year}
  {2017})}\BibitemShut {NoStop}%
\bibitem [{\citenamefont {Bisker}\ \emph {et~al.}(2017)\citenamefont {Bisker},
  \citenamefont {Polettini}, \citenamefont {Gingrich},\ and\ \citenamefont
  {Horowitz}}]{Bisker2017}%
  \BibitemOpen
  \bibfield  {author} {\bibinfo {author} {\bibfnamefont {G.}~\bibnamefont
  {Bisker}}, \bibinfo {author} {\bibfnamefont {M.}~\bibnamefont {Polettini}},
  \bibinfo {author} {\bibfnamefont {T.~R.}\ \bibnamefont {Gingrich}},\ and\
  \bibinfo {author} {\bibfnamefont {J.~M.}\ \bibnamefont {Horowitz}},\ }\href
  {https://doi.org/10.1088/1742-5468/aa8c0d} {\bibfield  {journal} {\bibinfo
  {journal} {Journal of Statistical Mechanics: Theory and Experiment}\ }\textbf
  {\bibinfo {volume} {2017}},\ \bibinfo {pages} {093210} (\bibinfo {year}
  {2017})}\BibitemShut {NoStop}%
\bibitem [{\citenamefont {Harunari}\ \emph {et~al.}(2022)\citenamefont
  {Harunari}, \citenamefont {Dutta}, \citenamefont {Polettini},\ and\
  \citenamefont {Rold\'an}}]{Harunari2022}%
  \BibitemOpen
  \bibfield  {author} {\bibinfo {author} {\bibfnamefont {P.~E.}\ \bibnamefont
  {Harunari}}, \bibinfo {author} {\bibfnamefont {A.}~\bibnamefont {Dutta}},
  \bibinfo {author} {\bibfnamefont {M.}~\bibnamefont {Polettini}},\ and\
  \bibinfo {author} {\bibfnamefont {E.}~\bibnamefont {Rold\'an}},\ }\href
  {https://doi.org/10.1103/PhysRevX.12.041026} {\bibfield  {journal} {\bibinfo
  {journal} {Phys. Rev. X}\ }\textbf {\bibinfo {volume} {12}},\ \bibinfo
  {pages} {041026} (\bibinfo {year} {2022})}\BibitemShut {NoStop}%
\bibitem [{\citenamefont {van~der Meer}\ \emph {et~al.}(2022)\citenamefont
  {van~der Meer}, \citenamefont {Ertel},\ and\ \citenamefont
  {Seifert}}]{vanderMeer2022}%
  \BibitemOpen
  \bibfield  {author} {\bibinfo {author} {\bibfnamefont {J.}~\bibnamefont
  {van~der Meer}}, \bibinfo {author} {\bibfnamefont {B.}~\bibnamefont
  {Ertel}},\ and\ \bibinfo {author} {\bibfnamefont {U.}~\bibnamefont
  {Seifert}},\ }\href {https://doi.org/10.1103/PhysRevX.12.031025} {\bibfield
  {journal} {\bibinfo  {journal} {Phys. Rev. X}\ }\textbf {\bibinfo {volume}
  {12}},\ \bibinfo {pages} {031025} (\bibinfo {year} {2022})}\BibitemShut
  {NoStop}%
\bibitem [{\citenamefont {Ertel}\ and\ \citenamefont
  {Seifert}(2024)}]{Ertel2024}%
  \BibitemOpen
  \bibfield  {author} {\bibinfo {author} {\bibfnamefont {B.}~\bibnamefont
  {Ertel}}\ and\ \bibinfo {author} {\bibfnamefont {U.}~\bibnamefont
  {Seifert}},\ }\href {https://doi.org/10.1103/PhysRevE.109.054109} {\bibfield
  {journal} {\bibinfo  {journal} {Phys. Rev. E}\ }\textbf {\bibinfo {volume}
  {109}},\ \bibinfo {pages} {054109} (\bibinfo {year} {2024})}\BibitemShut
  {NoStop}%
\bibitem [{\citenamefont {Ehrich}(2021)}]{Ehrich2021}%
  \BibitemOpen
  \bibfield  {author} {\bibinfo {author} {\bibfnamefont {J.}~\bibnamefont
  {Ehrich}},\ }\href {https://doi.org/10.1088/1742-5468/ac150e} {\bibfield
  {journal} {\bibinfo  {journal} {J. Stat. Mech.}\ }\textbf {\bibinfo {volume}
  {2021}},\ \bibinfo {pages} {083214} (\bibinfo {year} {2021})}\BibitemShut
  {NoStop}%
\bibitem [{\citenamefont {Dechant}(2018)}]{Dechant2018}%
  \BibitemOpen
  \bibfield  {author} {\bibinfo {author} {\bibfnamefont {A.}~\bibnamefont
  {Dechant}},\ }\href {https://doi.org/10.1088/1751-8121/aaf3ff} {\bibfield
  {journal} {\bibinfo  {journal} {J. Phys. A: Math. Theor.}\ }\textbf {\bibinfo
  {volume} {52}},\ \bibinfo {pages} {035001} (\bibinfo {year}
  {2018})}\BibitemShut {NoStop}%
\bibitem [{\citenamefont {Vu}\ \emph {et~al.}(2020)\citenamefont {Vu},
  \citenamefont {Vo},\ and\ \citenamefont {Hasegawa}}]{VVH2020}%
  \BibitemOpen
  \bibfield  {author} {\bibinfo {author} {\bibfnamefont {T.~V.}\ \bibnamefont
  {Vu}}, \bibinfo {author} {\bibfnamefont {V.~T.}\ \bibnamefont {Vo}},\ and\
  \bibinfo {author} {\bibfnamefont {Y.}~\bibnamefont {Hasegawa}},\ }\href
  {https://doi.org/10.1103/PhysRevE.101.042138} {\bibfield  {journal} {\bibinfo
   {journal} {Phys. Rev. E}\ }\textbf {\bibinfo {volume} {101}},\ \bibinfo
  {pages} {042138} (\bibinfo {year} {2020})}\BibitemShut {NoStop}%
\bibitem [{\citenamefont {Zia}\ and\ \citenamefont
  {Schmittmann}(2007)}]{ZS2007}%
  \BibitemOpen
  \bibfield  {author} {\bibinfo {author} {\bibfnamefont {R.~K.~P.}\
  \bibnamefont {Zia}}\ and\ \bibinfo {author} {\bibfnamefont {B.}~\bibnamefont
  {Schmittmann}},\ }\href {https://doi.org/10.1088/1742-5468/2007/07/P07012}
  {\bibfield  {journal} {\bibinfo  {journal} {Journal of Statistical Mechanics:
  Theory and Experiment}\ }\textbf {\bibinfo {volume} {2007}},\ \bibinfo
  {pages} {P07012} (\bibinfo {year} {2007})}\BibitemShut {NoStop}%
\bibitem [{Sup()}]{SuppMat}%
  \BibitemOpen
  \href@noop {} {}\bibinfo {note} {See Supplemental Material for proofs of the
  contraction, completion, screening, and cycle-observability results,
  simulation conventions, and finite-sample tests.}\BibitemShut {Stop}%
\bibitem [{\citenamefont {Gingrich}\ \emph {et~al.}(2017)\citenamefont
  {Gingrich}, \citenamefont {Rotskoff},\ and\ \citenamefont
  {Horowitz}}]{GRH2017}%
  \BibitemOpen
  \bibfield  {author} {\bibinfo {author} {\bibfnamefont {T.~R.}\ \bibnamefont
  {Gingrich}}, \bibinfo {author} {\bibfnamefont {G.~M.}\ \bibnamefont
  {Rotskoff}},\ and\ \bibinfo {author} {\bibfnamefont {J.~M.}\ \bibnamefont
  {Horowitz}},\ }\href {https://doi.org/10.1088/1751-8121/aa672f} {\bibfield
  {journal} {\bibinfo  {journal} {J. Phys. A: Math. Theor.}\ }\textbf {\bibinfo
  {volume} {50}},\ \bibinfo {pages} {184004} (\bibinfo {year}
  {2017})}\BibitemShut {NoStop}%
\bibitem [{\citenamefont {Liepelt}\ and\ \citenamefont
  {Lipowsky}(2007)}]{LL2007}%
  \BibitemOpen
  \bibfield  {author} {\bibinfo {author} {\bibfnamefont {S.}~\bibnamefont
  {Liepelt}}\ and\ \bibinfo {author} {\bibfnamefont {R.}~\bibnamefont
  {Lipowsky}},\ }\href {https://doi.org/10.1103/PhysRevLett.98.258102}
  {\bibfield  {journal} {\bibinfo  {journal} {Phys. Rev. Lett.}\ }\textbf
  {\bibinfo {volume} {98}},\ \bibinfo {pages} {258102} (\bibinfo {year}
  {2007})}\BibitemShut {NoStop}%
\bibitem [{\citenamefont {Liepelt}\ and\ \citenamefont
  {Lipowsky}(2009)}]{LL2009}%
  \BibitemOpen
  \bibfield  {author} {\bibinfo {author} {\bibfnamefont {S.}~\bibnamefont
  {Liepelt}}\ and\ \bibinfo {author} {\bibfnamefont {R.}~\bibnamefont
  {Lipowsky}},\ }\href {https://doi.org/10.1103/PhysRevE.79.011917} {\bibfield
  {journal} {\bibinfo  {journal} {Phys. Rev. E}\ }\textbf {\bibinfo {volume}
  {79}},\ \bibinfo {pages} {011917} (\bibinfo {year} {2009})}\BibitemShut
  {NoStop}%
\end{thebibliography}%


\begin{thebibliography}{6}%
\makeatletter
\providecommand \@ifxundefined [1]{%
 \@ifx{#1\undefined}
}%
\providecommand \@ifnum [1]{%
 \ifnum #1\expandafter \@firstoftwo
 \else \expandafter \@secondoftwo
 \fi
}%
\providecommand \@ifx [1]{%
 \ifx #1\expandafter \@firstoftwo
 \else \expandafter \@secondoftwo
 \fi
}%
\providecommand \natexlab [1]{#1}%
\providecommand \enquote  [1]{``#1''}%
\providecommand \bibnamefont  [1]{#1}%
\providecommand \bibfnamefont [1]{#1}%
\providecommand \citenamefont [1]{#1}%
\providecommand \href@noop [0]{\@secondoftwo}%
\providecommand \href [0]{\begingroup \@sanitize@url \@href}%
\providecommand \@href[1]{\@@startlink{#1}\@@href}%
\providecommand \@@href[1]{\endgroup#1\@@endlink}%
\providecommand \@sanitize@url [0]{\catcode `\\12\catcode `\$12\catcode
  `\&12\catcode `\#12\catcode `\^12\catcode `\_12\catcode `\%12\relax}%
\providecommand \@@startlink[1]{}%
\providecommand \@@endlink[0]{}%
\providecommand \url  [0]{\begingroup\@sanitize@url \@url }%
\providecommand \@url [1]{\endgroup\@href {#1}{\urlprefix }}%
\providecommand \urlprefix  [0]{URL }%
\providecommand \Eprint [0]{\href }%
\providecommand \doibase [0]{https://doi.org/}%
\providecommand \selectlanguage [0]{\@gobble}%
\providecommand \bibinfo  [0]{\@secondoftwo}%
\providecommand \bibfield  [0]{\@secondoftwo}%
\providecommand \translation [1]{[#1]}%
\providecommand \BibitemOpen [0]{}%
\providecommand \bibitemStop [0]{}%
\providecommand \bibitemNoStop [0]{.\EOS\space}%
\providecommand \EOS [0]{\spacefactor3000\relax}%
\providecommand \BibitemShut  [1]{\csname bibitem#1\endcsname}%
\let\auto@bib@innerbib\@empty
\bibitem [{\citenamefont {Bertini}\ \emph {et~al.}(2015)\citenamefont
  {Bertini}, \citenamefont {Faggionato},\ and\ \citenamefont
  {Gabrielli}}]{BFG2015}%
  \BibitemOpen
  \bibfield  {author} {\bibinfo {author} {\bibfnamefont {L.}~\bibnamefont
  {Bertini}}, \bibinfo {author} {\bibfnamefont {A.}~\bibnamefont
  {Faggionato}},\ and\ \bibinfo {author} {\bibfnamefont {D.}~\bibnamefont
  {Gabrielli}},\ }in\ \href@noop {} {\emph {\bibinfo {booktitle} {Annales de
  l'IHP Probabilit{\'e}s et statistiques}}},\ Vol.~\bibinfo {volume} {51}\
  (\bibinfo {year} {2015})\ pp.\ \bibinfo {pages} {867--900}\BibitemShut
  {NoStop}%
\bibitem [{\citenamefont {Maes}\ and\ \citenamefont
  {Netočný}(2007)}]{MN2007}%
  \BibitemOpen
  \bibfield  {author} {\bibinfo {author} {\bibfnamefont {C.}~\bibnamefont
  {Maes}}\ and\ \bibinfo {author} {\bibfnamefont {K.}~\bibnamefont
  {Netočný}},\ }\href {https://doi.org/10.1016/j.crhy.2007.05.003} {\bibfield
   {journal} {\bibinfo  {journal} {Comptes Rendus Physique}\ }\textbf {\bibinfo
  {volume} {8}},\ \bibinfo {pages} {591} (\bibinfo {year} {2007})},\ \bibinfo
  {note} {work, dissipation, and fluctuations in nonequilibrium
  physics}\BibitemShut {NoStop}%
\bibitem [{\citenamefont {Gingrich}\ \emph {et~al.}(2016)\citenamefont
  {Gingrich}, \citenamefont {Horowitz}, \citenamefont {Perunov},\ and\
  \citenamefont {England}}]{Gingrich2016}%
  \BibitemOpen
  \bibfield  {author} {\bibinfo {author} {\bibfnamefont {T.~R.}\ \bibnamefont
  {Gingrich}}, \bibinfo {author} {\bibfnamefont {J.~M.}\ \bibnamefont
  {Horowitz}}, \bibinfo {author} {\bibfnamefont {N.}~\bibnamefont {Perunov}},\
  and\ \bibinfo {author} {\bibfnamefont {J.~L.}\ \bibnamefont {England}},\
  }\href {https://doi.org/10.1103/PhysRevLett.116.120601} {\bibfield  {journal}
  {\bibinfo  {journal} {Phys. Rev. Lett.}\ }\textbf {\bibinfo {volume} {116}},\
  \bibinfo {pages} {120601} (\bibinfo {year} {2016})}\BibitemShut {NoStop}%
\bibitem [{\citenamefont {Polettini}\ \emph {et~al.}(2016)\citenamefont
  {Polettini}, \citenamefont {Lazarescu},\ and\ \citenamefont
  {Esposito}}]{PLE2016}%
  \BibitemOpen
  \bibfield  {author} {\bibinfo {author} {\bibfnamefont {M.}~\bibnamefont
  {Polettini}}, \bibinfo {author} {\bibfnamefont {A.}~\bibnamefont
  {Lazarescu}},\ and\ \bibinfo {author} {\bibfnamefont {M.}~\bibnamefont
  {Esposito}},\ }\href {https://doi.org/10.1103/PhysRevE.94.052104} {\bibfield
  {journal} {\bibinfo  {journal} {Phys. Rev. E}\ }\textbf {\bibinfo {volume}
  {94}},\ \bibinfo {pages} {052104} (\bibinfo {year} {2016})}\BibitemShut
  {NoStop}%
\bibitem [{\citenamefont {Liepelt}\ and\ \citenamefont
  {Lipowsky}(2007)}]{LL2007}%
  \BibitemOpen
  \bibfield  {author} {\bibinfo {author} {\bibfnamefont {S.}~\bibnamefont
  {Liepelt}}\ and\ \bibinfo {author} {\bibfnamefont {R.}~\bibnamefont
  {Lipowsky}},\ }\href {https://doi.org/10.1103/PhysRevLett.98.258102}
  {\bibfield  {journal} {\bibinfo  {journal} {Phys. Rev. Lett.}\ }\textbf
  {\bibinfo {volume} {98}},\ \bibinfo {pages} {258102} (\bibinfo {year}
  {2007})}\BibitemShut {NoStop}%
\bibitem [{\citenamefont {Liepelt}\ and\ \citenamefont
  {Lipowsky}(2009)}]{LL2009}%
  \BibitemOpen
  \bibfield  {author} {\bibinfo {author} {\bibfnamefont {S.}~\bibnamefont
  {Liepelt}}\ and\ \bibinfo {author} {\bibfnamefont {R.}~\bibnamefont
  {Lipowsky}},\ }\href {https://doi.org/10.1103/PhysRevE.79.011917} {\bibfield
  {journal} {\bibinfo  {journal} {Phys. Rev. E}\ }\textbf {\bibinfo {volume}
  {79}},\ \bibinfo {pages} {011917} (\bibinfo {year} {2009})}\BibitemShut
  {NoStop}%
\end{thebibliography}%

\end{document}


\title{Supplemental Material for ``Certifying Hidden Dissipation
       from Observed Current Fluctuations''}

\author{Ahmed Roman}
\affiliation{Department of Medical Oncology, Dana-Farber Cancer Institute,
Boston, Massachusetts 02215, USA}
\affiliation{Broad Institute of MIT and Harvard,
Cambridge, Massachusetts 02142, USA}
\affiliation{Harvard Medical School, Boston, Massachusetts 02115, USA}

\date{September 11, 2026}

\maketitle

This Supplemental Material provides complete, self-contained
derivations for all results stated in the main text. Equations are
numbered (S1), (S2), \ldots\ for cross-reference from the main text.
Theorems~S1--S6 contain the full statements that the main text
replaces with narrative summaries; the sections below give the proofs.

\setcounter{equation}{0}
\renewcommand{\theequation}{S\arabic{equation}}
\renewcommand{\thefigure}{S\arabic{figure}}

\section{Level-2.5 quadratic contraction}
\label{sec:contraction}

As in the main text, the chain is finite and irreducible, with positive
rates in both directions on every included edge.
At long observation time $t$, the joint fluctuations of the empirical
density $\hat\rho$ and the empirical flow $\hat\phi_{ij}$ satisfy a
large-deviation principle with speed $t$, so probabilities scale
exponentially as $\exp[-tI(\rho,\phi)]$, where
$I[\rho,\phi]$ is the level-2.5 functional
\cite{BFG2015,MN2007}. Restricting to stationary continuity
$\sum_j(\phi_{ij}-\phi_{ji})=0$ and expanding to second order around
the typical point $\rho=p$, $\phi_{ij}=q_{ij}$ gives
\begin{equation}
I^{(2)}(\delta\rho,\delta\phi)
= \tfrac12 \sum_{i,j}
  \frac{(\delta\phi_{ij} - w_{ij}\,\delta\rho_i)^2}{q_{ij}},
\label{eq:S-level25}
\end{equation}
with the sum over directed pairs $(i,j)$ such that $w_{ij}>0$.

Per undirected edge $e=\{i,j\}$ with chosen orientation $i\to j$,
define edge-current and edge-traffic fluctuations
$\delta K_e = \delta\phi_{ij}-\delta\phi_{ji}$ and
$\delta T_e = \delta\phi_{ij}+\delta\phi_{ji}$, and the
density-gradient operator
$(G\,\delta\rho)_e \equiv w_{ij}\,\delta\rho_i - w_{ji}\,\delta\rho_j$.
Set $x=\delta K_e-(G\delta\rho)_e$ and
$y=\delta T_e-w_{ij}\delta\rho_i-w_{ji}\delta\rho_j$.
The directed-pair contribution is
$(x+y)^2/(8q_{ij})+(y-x)^2/(8q_{ji})$.
Traffic is unconstrained but coupled to current; minimizing over $y$
gives $y=(K_e/T_e)x$ and the edge cost
\begin{equation}
\frac{(\delta K_e - (G\,\delta\rho)_e)^2}{2T_e},
\end{equation}
where $T_e = q_{ij}+q_{ji}$ is the stationary symmetric activity
(traffic) on edge $e$. Summing over edges:
\begin{equation}
I^{(2)}(\delta K,\delta\rho)
= \tfrac12\,(\delta K - G\,\delta\rho)^\top
  D_T^{-1}\,(\delta K - G\,\delta\rho),
\label{eq:S-I-after-T}
\end{equation}
with $D_T = \mathrm{diag}(T_e)$.

\subsection{Contraction over normalized density}

Density fluctuations satisfy $\mathbf{1}^\top\delta\rho = 0$
(probability conservation). Let $R\in\mathbb{R}^{N\times(N-1)}$ be a
basis for $\mathbf{1}^\perp$, so $\delta\rho = R\xi$ with
$\xi\in\mathbb{R}^{N-1}$. Define $\tilde G \equiv GR$.
The matrix $\tilde G^\top D_T^{-1}\tilde G$ is positive definite.
Indeed, if $\tilde G\xi=0$, then $x=R\xi$ satisfies
$\mathbf 1^\top x=0$ and $Gx=0$. Summing the edge fluxes at each
state gives $L^\top x=0$, where $L$ is the Markov generator
[see \eqref{eq:S-div-G}]. Irreducibility implies $x=cp$; normalization
gives $c=0$, and full column rank of $R$ gives $\xi=0$.
Thus $\tilde G$ has trivial kernel, and $D_T^{-1}\succ0$ establishes
the claim. Minimizing
\eqref{eq:S-I-after-T} over $\xi$ at fixed $\delta K$ gives the
unique optimizing density shift
\begin{equation}
\xi^* = (\tilde G^\top D_T^{-1}\tilde G)^{-1}\,
        \tilde G^\top D_T^{-1}\,\delta K,
\label{eq:S-xistar}
\end{equation}
and the contracted rate function on edge currents alone:
\begin{equation}
I^{(2)}_{\rm cyc}(\delta K)
= \tfrac12\,\delta K^\top M_T\,\delta K,
\label{eq:S-cyc-rate}
\end{equation}
with the density-contracted traffic metric
\begin{equation}
M_T = D_T^{-1}
    - D_T^{-1}\tilde G\,
      (\tilde G^\top D_T^{-1}\tilde G)^+\,
      \tilde G^\top D_T^{-1}.
\label{eq:S-MT}
\end{equation}
The pseudoinverse notation retained for this density block coincides
with its ordinary inverse.
The whitened-projector representation \eqref{eq:S-MT-projector} gives
$M_T \preceq D_T^{-1}$ in the positive-semidefinite sense. Equality
of the two quadratic forms on the entire cycle subspace,
\begin{equation}
J^\top M_TJ = J^\top D_T^{-1}J
\qquad\text{for every }J\in\mathcal C,
\label{eq:S-global-equality-form}
\end{equation}
holds if and only if
\begin{equation}
C^\top D_T^{-1}\tilde G = 0.
\label{eq:S-global-orthogonality}
\end{equation}
Equivalently, $M_TC = D_T^{-1}C$. A stronger current-specific and
global characterization is proved in the
\hyperref[sec:screening]{screening-orthogonality section}.

\section{Theorem S1: Density-contracted traffic metric}
\label{sec:thmS1}

\textbf{Theorem S1.}
\emph{The Hessian of the level-2.5 rate function, contracted over
normalized empirical-density fluctuations and traffic fluctuations,
is the density-contracted traffic metric $M_T$ defined in
\eqref{eq:S-MT}. The asymptotic CLT covariance of empirical edge
currents, restricted to the cycle subspace
$\mathcal{C} = \mathrm{col}(C)$, is}
\begin{equation}
\Sigma_K = C\,(C^\top M_T\,C)^{-1}\,C^\top.
\label{eq:S-Sigma}
\end{equation}
\emph{$\Sigma_K$ has rank $r = |\mathcal{E}|-N+1$ and vanishes on
the divergence-carrying complement of $\mathcal{C}$.}

\medskip
\emph{Proof.}
The contracted rate function \eqref{eq:S-cyc-rate} is the Gaussian
rate function for edge-current fluctuations after the traffic and
density sectors have been optimized out. Restricting to
divergence-free currents $\delta K = C\,\delta\alpha$ with
$\delta\alpha\in\mathbb{R}^r$, the rate function becomes
\begin{equation}
I^{(2)}_{\rm cyc}(C\,\delta\alpha)
= \tfrac12\,\delta\alpha^\top (C^\top M_T\,C)\,\delta\alpha.
\label{eq:S-cycle-gauss}
\end{equation}
The matrix $C^\top M_T\,C$ is positive definite for every finite
irreducible CTMC. To see this, first write
\begin{equation}
\begin{aligned}
M_T
&=D_T^{-1/2}\bigl[I-A(A^\top A)^+A^\top\bigr]D_T^{-1/2},\\
A&=D_T^{-1/2}\tilde G.
\end{aligned}
\label{eq:S-MT-projector}
\end{equation}
The bracket is the orthogonal projector onto
$\operatorname{col}(A)^\perp$. Hence $M_T$ is positive
semidefinite and
\begin{equation}
\ker M_T=\operatorname{col}(\tilde G).
\label{eq:S-kernel-MT}
\end{equation}
It remains to show that
\begin{equation}
\mathcal C\cap\operatorname{col}(\tilde G)=\{0\}.
\label{eq:S-cycle-density-intersection}
\end{equation}
Let $J=\tilde G\xi=GR\xi\in\mathcal C$, and set $x=R\xi$. Then
$\mathbf 1^\top x=0$ because the columns of $R$ span normalized
density perturbations. Let $L$ be the generator with
$L_{ij}=w_{ij}$ for $i\ne j$ and $L_{ii}=-\sum_{j\ne i}w_{ij}$, and
let $\partial$ be the oriented incidence matrix, so
$\mathcal C=\ker\partial$. For an oriented edge $e:i\to j$,
$(Gx)_e=w_{ij}x_i-w_{ji}x_j$, and with the incidence convention of
the main text,
\begin{equation}
(\partial Gx)_i
=\sum_j\bigl(w_{ji}x_j-w_{ij}x_i\bigr)
=(L^\top x)_i .
\label{eq:S-div-G}
\end{equation}
The opposite incidence convention changes only the overall sign. Since
$J\in\ker\partial$, Eq.~\eqref{eq:S-div-G} gives $L^\top x=0$.
Irreducibility implies $\ker L^\top=\operatorname{span}\{p\}$, where
$p$ is the normalized stationary distribution. Thus $x=cp$ for some
constant $c$. But $0=\mathbf 1^\top x=c\,\mathbf 1^\top p=c$, so
$x=0$ and $J=0$. This proves
\eqref{eq:S-cycle-density-intersection}. Therefore any nonzero
$C\delta\alpha$ has strictly positive $M_T$-cost, and
$C^\top M_T C\succ0$. The Gaussian inverse is the limiting covariance
of $\sqrt t\,\delta\alpha$, namely $(C^\top M_T\,C)^{-1}$.
Pushing back via $\delta K=C\,\delta\alpha$ yields
\eqref{eq:S-Sigma}, with
$\Sigma_K=\lim_{t\to\infty}t\,\mathrm{Cov}(\hat K)$ and
$\sqrt t(\hat K-K)\Rightarrow\mathcal N(0,\Sigma_K)$.
\hfill$\square$

\section{Theorem S2: Optimized covariance pseudoinverse}
\label{sec:thmS2}

\textbf{Theorem S2.}
\emph{For any vector $u$ with $u^\top\Sigma_Ku>0$, the
scalar TUR \cite{Gingrich2016} states
$\sigma \ge 2(u^\top K)^2/(u^\top\Sigma_K u)$. The supremum over
$u$ satisfies}
\begin{equation}
K^\top\Sigma_K^+ K
= \sup_{u^\top\Sigma_Ku>0}\frac{(u^\top K)^2}{u^\top\Sigma_K u}
= K^\top M_T K
\le K^\top D_T^{-1} K,
\label{eq:S-envelope}
\end{equation}
\emph{giving the optimized envelope
$\sigma \ge 2\,K^\top\Sigma_K^+ K = 2\,K^\top M_T K$.}

\medskip
\emph{Proof.}
The Rayleigh-quotient identity
$\sup_{u^\top\Sigma_Ku>0}(u^\top K)^2/(u^\top\Sigma_K u) = K^\top\Sigma_K^+ K$
holds for any positive-semidefinite $\Sigma_K$ when $K$ lies in its
column space (with value zero by convention if $\Sigma_K=0$).
Since $K = C\alpha$ is divergence-free and $\Sigma_K$
has column space $\mathcal{C} = \mathrm{col}(C)$, the condition is
satisfied. Substituting \eqref{eq:S-Sigma}:
\begin{align}
K^\top\Sigma_K^+ K
&= \alpha^\top C^\top\bigl[C(C^\top M_T C)^{-1}C^\top\bigr]^+
   C\alpha \nonumber\\
&= \alpha^\top (C^\top M_T C)\,\alpha \nonumber\\
&= K^\top M_T K,
\label{eq:S-pseudoinv-calc}
\end{align}
where the second line uses the pseudoinverse identity
$(C A^{-1} C^\top)^+ = C^{+\top} A\, C^+$ for full-column-rank $C$
and positive-definite $A = C^\top M_T C$. The inequality
$K^\top M_T K \le K^\top D_T^{-1} K$ follows from
$M_T \preceq D_T^{-1}$.
\hfill$\square$

\section{Pseudoinverse identity in edge space}
\label{sec:pseudoinv}

The covariance $\Sigma_K = C(C^\top M_T C)^{-1}C^\top$ has rank $r$
and lives on $\mathcal{C} = \mathrm{col}(C)$. Its pseudoinverse
satisfies the following matrix identity on the entire edge space:
\begin{equation}
\Sigma_K^+
= C^{+\top}(C^\top M_T C)\,C^+
= P_{\mathcal{C}}\,M_T\,P_{\mathcal{C}},
\label{eq:S-pinv-cycle}
\end{equation}
where $P_{\mathcal{C}} = C(C^\top C)^{-1}C^\top$ is the orthogonal
projection onto $\mathcal{C}$ and
$C^+ = (C^\top C)^{-1}C^\top$ is the left pseudoinverse of
full-column-rank $C$.

\emph{Derivation.}
Write $\Sigma_K = C A^{-1} C^\top$ with $A = C^\top M_T C$
positive definite. The pseudoinverse of a rank-$r$ matrix of the
form $C A^{-1} C^\top$ with full-column-rank $C$ is
\begin{equation}
(C A^{-1} C^\top)^+
= C(C^\top C)^{-1}\,A\,(C^\top C)^{-1}C^\top.
\label{eq:S-pinv-formula}
\end{equation}
Substituting $A = C^\top M_T C$:
\begin{align}
\Sigma_K^+
&= C(C^\top C)^{-1}(C^\top M_T C)(C^\top C)^{-1}C^\top
  \nonumber\\
&= P_{\mathcal{C}}\,M_T\,P_{\mathcal{C}}.
\label{eq:S-pinv-expand}
\end{align}
For any $K\in\mathcal{C}$, therefore
$K^\top\Sigma_K^+ K = K^\top M_T K$, confirming
Theorem~S2.
\hfill$\square$

\section{Theorem S3: Minimum-energy completion}
\label{sec:projection}

\textbf{Theorem S3 (Minimum-energy completion).}
\emph{Let $\Sigma = C(C^\top M_T C)^{-1}C^\top$ be the induced
covariance on $\mathbb{R}^{|\mathcal{E}|}$, and for an observed-edge
projection $P_{\mathcal{O}}$ let $\Sigma_{\mathcal{O}} =
P_{\mathcal{O}}\,\Sigma\,P_{\mathcal{O}}^\top$. Then the observed
covariance pseudoinverse satisfies the Gaussian projection identity}
\begin{equation}
K_{\mathcal{O}}^\top\Sigma_{\mathcal{O}}^+ K_{\mathcal{O}}
= \min_{\substack{J\in\mathcal{C}\\
                   P_{\mathcal{O}}J = K_{\mathcal{O}}}}
  J^\top M_T\,J.
\label{eq:S-projection}
\end{equation}
This is the boxed minimum-energy completion identity of the main text.

\emph{Proof.}
Write $J = C\beta$ for $J\in\mathcal{C}$, and let
$\tilde P = P_{\mathcal{O}}C$ so that the constraint becomes
$\tilde P\,\beta = K_{\mathcal{O}}$. Set
$A=C^\top M_T C$, which is positive definite by Theorem~S1. With
$z=A^{1/2}\beta$ and
\begin{equation}
B\equiv \tilde P A^{-1/2},
\end{equation}
the constrained problem becomes
\begin{equation}
\min_{Bz=K_{\mathcal O}}\|z\|^2 .
\label{eq:S-min-norm}
\end{equation}
The constraint is feasible because $K\in\mathcal C$ and
$K_{\mathcal O}=P_{\mathcal O}K$. For any feasible right-hand side
$y\in\operatorname{ran}(B)$, the minimum-norm solution is
$z=B^+y$, hence
\begin{equation}
\min_{Bz=y}\|z\|^2
=y^\top (B^+)^\top B^+y
=y^\top (BB^\top)^+y.
\label{eq:S-min-norm-pinv}
\end{equation}
The last equality follows immediately from a singular-value
decomposition of $B$.
Here
\begin{equation}
BB^\top
=\tilde P A^{-1}\tilde P^\top
=P_{\mathcal O}C(C^\top M_T C)^{-1}C^\top P_{\mathcal O}^\top
=\Sigma_{\mathcal O}.
\end{equation}
Taking $y=K_{\mathcal O}$ gives \eqref{eq:S-projection}, including
the case where $\Sigma_{\mathcal O}$ is rank deficient.
\hfill$\square$

\medskip
\textbf{Proposition (rank criterion for cycle observability).}
\emph{With $r=\dim\mathcal C$, the observed-edge projection
$P_{\mathcal O}|_{\mathcal C}$ is injective if and only if the exact
long-time observed covariance satisfies}
\begin{equation}
\operatorname{rank}\Sigma_{\mathcal O}=r.
\label{eq:S-rank-criterion}
\end{equation}
\emph{Equivalently, if the total cycle rank $r$ is known
independently, cycle observability can be certified from the rank of
the observed-current covariance.}

\emph{Proof.}
Let $A=C^\top M_T C$. By Theorem~S1, $A$ is positive definite. With
$X=P_{\mathcal O}C$,
\begin{equation}
\Sigma_{\mathcal O}=X A^{-1}X^\top .
\end{equation}
Choose an invertible square root $S=A^{-1/2}$. Then
\begin{equation}
\Sigma_{\mathcal O}=(XS)(XS)^\top ,
\end{equation}
so
\begin{equation}
\operatorname{rank}\Sigma_{\mathcal O}
=\operatorname{rank}(XS)
=\operatorname{rank}(X)
=\operatorname{rank}(P_{\mathcal O}C).
\end{equation}
Because $C$ has full column rank, $P_{\mathcal O}|_{\mathcal C}$ is
injective exactly when $P_{\mathcal O}C$ has rank $r$. This proves
\eqref{eq:S-rank-criterion}. If the hidden topology and the total
cycle rank are both unknown, the bound remains valid, but exact
recovery of the full cycle cost cannot be certified from visible
statistics alone.
\hfill$\square$

\section{Screening orthogonality and $\gamma_G$}
\label{sec:screening}

The screening fraction is current-specific. For a nonzero cycle
current $K\in\mathcal C$, define
\begin{align}
\gamma_G(K)
&= \frac{K^\top(D_T^{-1}-M_T)K}{K^\top D_T^{-1} K}
\nonumber\\
&= \frac{K^\top D_T^{-1}\tilde G\,
        (\tilde G^\top D_T^{-1}\tilde G)^+\,
        \tilde G^\top D_T^{-1} K}
       {K^\top D_T^{-1} K}.
\label{eq:S-gamma}
\end{align}
Equivalently, with
\begin{align}
Q_G
&\equiv D_T^{-1}-M_T
\nonumber\\
&=D_T^{-1}\tilde G
  (\tilde G^\top D_T^{-1}\tilde G)^+
  \tilde G^\top D_T^{-1},
\label{eq:S-QG}
\end{align}
one has
\begin{equation}
\gamma_G(K)=\frac{K^\top Q_GK}{K^\top D_T^{-1}K}.
\label{eq:S-gamma-QG}
\end{equation}

\textbf{Proposition S7 (current-specific and global screening criteria).}
For a fixed nonzero current $K$, the following statements are
equivalent:
\begin{align}
\gamma_G(K)=0
&\iff
\tilde G^\top D_T^{-1}K=0,
\label{eq:S-fixed-screening-a}\\
&\iff
K\perp \operatorname{col}(D_T^{-1}\tilde G).
\label{eq:S-fixed-screening-b}
\end{align}
The orthogonality in \eqref{eq:S-fixed-screening-b} is in the
standard edge-space inner product.
In particular, if $K=C\alpha$, then
\begin{equation}
\gamma_G(C\alpha)=0
\iff
\tilde G^\top D_T^{-1}C\alpha=0.
\label{eq:S-fixed-screening-cycle}
\end{equation}
Thus, vanishing screening for one current only requires
\begin{equation}
\alpha\in\ker\!\left(\tilde G^\top D_T^{-1}C\right).
\label{eq:S-fixed-kernel}
\end{equation}

The following stronger statements are equivalent:
\begin{align}
&\gamma_G(C\alpha)=0
\quad
\text{for every nonzero }\alpha\in\mathbb R^r,
\label{eq:S-global-screening-a}\\
&\tilde G^\top D_T^{-1}C=0,
\label{eq:S-global-screening-b}\\
&C^\top D_T^{-1}\tilde G=0,
\label{eq:S-global-screening-c}\\
&(D_T^{-1}-M_T)C=0,
\label{eq:S-global-screening-d}\\
&M_TC=D_T^{-1}C,
\label{eq:S-global-screening-e}\\
&C^\top M_TC=C^\top D_T^{-1}C.
\label{eq:S-global-screening-f}
\end{align}
These conditions express absence of density screening on the
\emph{entire} cycle subspace.

\medskip
\emph{Proof.}
Set
\begin{equation}
A\equiv \tilde G^\top D_T^{-1}\tilde G,
\qquad
y_K\equiv \tilde G^\top D_T^{-1}K.
\end{equation}
Then
\begin{equation}
K^\top Q_GK = y_K^\top A^+y_K.
\end{equation}
Because $D_T^{-1}$ is positive definite,
\begin{equation}
x^\top Ax
= \left\|D_T^{-1/2}\tilde Gx\right\|^2,
\end{equation}
and therefore $\ker(A)=\ker(\tilde G)$. Since $A$ is symmetric,
\begin{equation}
\operatorname{ran}(A)
=\ker(A)^\perp
=\ker(\tilde G)^\perp
=\operatorname{ran}(\tilde G^\top).
\end{equation}
Moreover,
$y_K=\tilde G^\top D_T^{-1}K\in\operatorname{ran}(\tilde G^\top)
=\operatorname{ran}(A)$. The pseudoinverse $A^+$ is positive
definite on $\operatorname{ran}(A)$, hence
\begin{equation}
y_K^\top A^+y_K=0
\iff
y_K=0.
\end{equation}
This proves
$\gamma_G(K)=0\iff \tilde G^\top D_T^{-1}K=0$. The orthogonality
formulation follows from
\begin{equation}
\tilde G^\top D_T^{-1}K=0
\iff
K^\top D_T^{-1}\tilde G=0.
\end{equation}
For $K=C\alpha$, this becomes
\begin{equation}
\gamma_G(C\alpha)=0
\iff
\tilde G^\top D_T^{-1}C\alpha=0.
\end{equation}
It holds for every $\alpha$ if and only if
\begin{equation}
\tilde G^\top D_T^{-1}C=0,
\end{equation}
which is equivalent by transposition to
$C^\top D_T^{-1}\tilde G=0$.

Finally,
\begin{equation}
(D_T^{-1}-M_T)C
=D_T^{-1}\tilde G A^+\tilde G^\top D_T^{-1}C,
\end{equation}
so the matrix orthogonality condition implies
$(D_T^{-1}-M_T)C=0$ and hence $M_TC=D_T^{-1}C$. Conversely, equality
of the quadratic forms on the cycle subspace gives
\begin{equation}
0=C^\top(D_T^{-1}-M_T)C=B^\top A^+B,
\qquad
B\equiv\tilde G^\top D_T^{-1}C.
\end{equation}
Every column of $B$ lies in $\operatorname{ran}(A)$, where $A^+$ is
positive definite. Therefore $B^\top A^+B=0$ implies $B=0$. This
proves all equivalences.
\hfill$\square$

\medskip
\textbf{Unscreened cycle subspace.}
Let
\begin{equation}
B_G\equiv \tilde G^\top D_T^{-1}C.
\end{equation}
The cycle currents that experience no density screening form the
linear subspace
\begin{equation}
\mathcal U_G
\equiv
\{0\}\cup\{C\alpha:\alpha\ne0,\ \gamma_G(C\alpha)=0\}
= C\,\ker(B_G).
\label{eq:S-unscreened-subspace}
\end{equation}
Because $C$ has full column rank,
\begin{equation}
\dim\mathcal U_G
= r-\operatorname{rank}(B_G).
\label{eq:S-unscreened-dim}
\end{equation}
Consequently, the global orthogonality condition
$C^\top D_T^{-1}\tilde G=0$ is precisely the special case
$\mathcal U_G=\mathcal C$. In general, a proper subspace of cycle
currents may be unscreened even when the density and cycle sectors
are not globally orthogonal.

\medskip
\textbf{Cut/gradient formulation of global orthogonality.}
Let $B_{\mathrm{inc}}$ denote the oriented incidence matrix, so
$\mathcal C=\ker B_{\mathrm{inc}}$. Then
\begin{equation}
C^\top D_T^{-1}\tilde G=0
\iff
\operatorname{col}(D_T^{-1}\tilde G)
\subseteq
\operatorname{col}(B_{\mathrm{inc}}^\top).
\label{eq:S-cut-gradient}
\end{equation}
Indeed, $C^\top X=0$ if and only if every column of $X$ belongs to
$(\ker B_{\mathrm{inc}})^\perp
=\operatorname{col}(B_{\mathrm{inc}}^\top)$, and taking
$X=D_T^{-1}\tilde G$ gives the claim. Thus the precise global
condition is that the weighted density-gradient directions
$D_T^{-1}\tilde G$ lie in the cut/gradient subspace.

\medskip
\textbf{Proposition S8 (symmetry-protected absence of screening for
the stationary current).}
Suppose that a group $\Gamma$ acts transitively on the state space and
leaves the transition rates invariant:
\begin{equation}
w_{g(i)g(j)}=w_{ij}
\qquad
\text{for every }g\in\Gamma.
\end{equation}
For an irreducible chain, the stationary distribution is then uniform.
If the stationary current $K$ is nonzero, it satisfies
\begin{equation}
\gamma_G(K)=0.
\end{equation}
This statement concerns the physical stationary current $K$. It does
not, in general, imply $C^\top D_T^{-1}\tilde G=0$, which would
require vanishing screening for every cycle current.

\medskip
\emph{Proof.}
Write the stationary current and traffic on an unoriented edge
$\{i,j\}$ as
\begin{equation}
K_{ij}=p_iw_{ij}-p_jw_{ji},
\qquad
T_{ij}=p_iw_{ij}+p_jw_{ji},
\end{equation}
with $K_{ji}=-K_{ij}$ and $T_{ji}=T_{ij}$. Here $K_{ij}$ denotes the
signed current from $i$ to $j$, independently of the chosen orientation
of the edge. For an arbitrary density perturbation $\delta\rho$,
\begin{equation}
K^\top D_T^{-1}G\,\delta\rho
=\sum_i h_i\,\delta\rho_i,
\qquad
h_i=\sum_{j:\{i,j\}\in\mathcal E}
\frac{K_{ij}}{T_{ij}}\,w_{ij}.
\end{equation}
Rate invariance implies
$K_{g(i)g(j)}=K_{ij}$ and $T_{g(i)g(j)}=T_{ij}$, and therefore
$h_{g(i)}=h_i$. Because the group action is transitive, $h_i$ is
independent of $i$, so $h=c\mathbf 1$ for some scalar $c$.
Normalized density perturbations satisfy
$\mathbf 1^\top\delta\rho=0$, and therefore
\begin{equation}
K^\top D_T^{-1}G\,\delta\rho
=c\,\mathbf 1^\top\delta\rho
=0.
\end{equation}
Since every normalized perturbation can be written as
$\delta\rho=R\xi$, this gives
$K^\top D_T^{-1}\tilde G=0$, equivalently
$\tilde G^\top D_T^{-1}K=0$. By Proposition~S7,
$\gamma_G(K)=0$.
\hfill$\square$

\section{Proof of the hybrid bound and monotonicity}
\label{sec:hybrid}

For completeness, we reproduce the proof of the hybrid bound and its
monotonicity.

\textbf{Theorem S4 (Hybrid passive lower bound).}
The Schnakenberg edge entropy production rate decomposes as
$\sigma_e = 2K_e^2/T_e + R_e$ with nonnegative remainder
\begin{equation}
R_e = K_e\log\!\Bigl(\frac{T_e+K_e}{T_e-K_e}\Bigr)
    - 2\frac{K_e^2}{T_e} \ge 0.
\label{eq:S-remainder}
\end{equation}
Set $\eta_e=K_e/T_e$. Since $T_e>|K_e|$, one has $|\eta_e|<1$.
The function $f(\eta)=\mathrm{artanh}(\eta)-\eta$ vanishes at zero
and has derivative $f'(\eta)=\eta^2/(1-\eta^2)\ge0$, so it has
the same sign as $\eta$. Hence
$R_e=2T_e\eta_e[\mathrm{artanh}(\eta_e)-\eta_e]\ge0$
for either sign of the current, including $R_e=0$ at $K_e=0$.

For any observed subset $\mathcal{O}\subseteq\mathcal{E}$:
\begin{align}
\sigma &= 2K^\top D_T^{-1}K + \sum_{e\in\mathcal{E}} R_e
\nonumber\\
&\ge 2K^\top M_T K + \sum_{e\in\mathcal{E}} R_e
\nonumber\\
&\ge 2K^\top M_T K + \sum_{e\in\mathcal{O}} R_e
\nonumber\\
&\ge 2K_{\mathcal{O}}^\top\Sigma_{\mathcal{O}}^+ K_{\mathcal{O}}
   + \sum_{e\in\mathcal{O}} R_e
= \mathcal{B}_{\mathcal{O}}.
\label{eq:S-hybrid-chain}
\end{align}
The first inequality uses $M_T\preceq D_T^{-1}$; the second drops
$R_e\ge 0$ on hidden edges; the third applies the minimum-completion
identity (Theorem~S3).
\hfill$\square$

\textbf{Theorem S5 (Monotonicity of $\mathcal{B}_{\mathcal{O}}^{\max}$).}
For $\mathcal{O}\subseteq\mathcal{O}'$, the constraint
$P_{\mathcal{O}'}J = K_{\mathcal{O}'}$ implies
$P_{\mathcal{O}}J = K_{\mathcal{O}}$, so the feasible set in the
minimum-energy completion shrinks:
\begin{equation}
K_{\mathcal{O}}^\top\Sigma_{\mathcal{O}}^+ K_{\mathcal{O}}
\le K_{\mathcal{O}'}^\top\Sigma_{\mathcal{O}'}^+ K_{\mathcal{O}'}.
\label{eq:S-mono-cov}
\end{equation}
Adding $\sum_{\mathcal{O}}R_e \le \sum_{\mathcal{O}'}R_e$ and
$\sum_{\mathcal{O}}\sigma_e \le \sum_{\mathcal{O}'}\sigma_e$
gives monotonicity of
$\mathcal{B}_{\mathcal{O}}^{\max}
= \max\{\sum_{\mathcal{O}}\sigma_e,\,\mathcal{B}_{\mathcal{O}}\}$.
\hfill$\square$

\section{Cycle observability and strict hidden-edge advantage}
\label{sec:cycle-obs}

\textbf{Theorem S6 (Cycle observability).}
\emph{If $P_{\mathcal{O}}:\mathcal{C}\to\mathbb{R}^{|\mathcal{O}|}$
is injective, then the unique divergence-free current with
$P_{\mathcal{O}}J = K_{\mathcal{O}}$ is $J = K$ itself, and
$K_{\mathcal{O}}^\top\Sigma_{\mathcal{O}}^+ K_{\mathcal{O}}
= K^\top M_T K$.}

\medskip
\emph{Proof.}
Injectivity of $P_{\mathcal{O}}|_{\mathcal{C}}$ means the feasible
set $\{J\in\mathcal{C}: P_{\mathcal{O}}J = K_{\mathcal{O}}\}$ is a
singleton $\{K\}$. The minimum in Theorem~S3 is therefore achieved
at $J = K$, giving
$K_{\mathcal{O}}^\top\Sigma_{\mathcal{O}}^+ K_{\mathcal{O}}
= K^\top M_T K$.
\hfill$\square$

\textbf{Corollary (Strict hidden-edge advantage for an unscreened
physical current).}
Under the conditions of Theorem~S6, suppose that the actual stationary
current satisfies
\begin{equation}
\gamma_G(K)=0,
\end{equation}
and that at least one hidden edge $h\notin\mathcal O$ carries
$K_h\neq0$. Then
\begin{equation}
K_{\mathcal{O}}^\top\Sigma_{\mathcal{O}}^+ K_{\mathcal{O}}
= \sum_{e\in\mathcal{E}}\frac{K_e^2}{T_e}
> \sum_{e\in\mathcal{O}}\frac{K_e^2}{T_e}.
\label{eq:S-strict}
\end{equation}

\emph{Proof.}
From Theorem~S6,
$K_{\mathcal{O}}^\top\Sigma_{\mathcal{O}}^+ K_{\mathcal{O}}
= K^\top M_T K$. The current-specific condition
$\gamma_G(K)=0$ is equivalent to
$K^\top M_TK=K^\top D_T^{-1}K$, so
\begin{equation}
K_{\mathcal{O}}^\top\Sigma_{\mathcal{O}}^+ K_{\mathcal{O}}
= K^\top D_T^{-1}K
= \sum_{e\in\mathcal E}\frac{K_e^2}{T_e}.
\end{equation}
Since $h\notin\mathcal O$, $K_h\neq0$, and $T_h>0$,
\begin{equation}
\sum_{e\in\mathcal E}\frac{K_e^2}{T_e}
- \sum_{e\in\mathcal O}\frac{K_e^2}{T_e}
= \sum_{e\notin\mathcal O}\frac{K_e^2}{T_e}
\ge \frac{K_h^2}{T_h}
>0.
\end{equation}
This proves the strict inequality.
\hfill$\square$

\section{Relation to force-weighted conductances}
\label{sec:force}

The Schnakenberg edge entropy production uses the thermodynamic
force $F_e = \log(q_{ij}/q_{ji}) = \log[(T_e+K_e)/(T_e-K_e)]
= 2\,\mathrm{artanh}(\eta_e)$, so
$\sigma_e = K_e F_e = 2T_e\eta_e\,\mathrm{artanh}(\eta_e)$.
The force-weighted conductance is, using $K_e = \eta_e T_e$,
$F_e/(2K_e) = \mathrm{artanh}(\eta_e)/(\eta_e T_e)
= (1/T_e)\,\mathrm{artanh}(\eta_e)/\eta_e$,
with the ratio at $K_e=0$ defined by continuity. It admits the Taylor expansion
\begin{equation}
\frac{F_e}{2K_e}
= \frac{1}{T_e}\Bigl(1 + \frac{\eta_e^2}{3}
  + \frac{\eta_e^4}{5} + \cdots\Bigr).
\label{eq:S-force-expand}
\end{equation}
At leading order, the force conductance reduces to $1/T_e$, which is
the local metric entering $M_T$. The higher-order terms capture the
nonlinear Schnakenberg remainder $R_e$; specifically,
$R_e = 2T_e\eta_e[\mathrm{artanh}(\eta_e) - \eta_e]
= 2K_e^2[F_e/(2K_e) - 1/T_e]$, which is manifestly nonnegative.

\section{Computing the bound from data}
\label{sec:algorithm}

Every quantity in the operational bound
$\widehat{\mathcal{B}}_{\mathcal{O}}^{\max}
= \max\{\sum_{e\in\mathcal{O}}\hat\sigma_e,\;
2\hat K_{\mathcal{O}}^\top\hat\Sigma_{\mathcal{O}}^+\hat K_{\mathcal{O}}
+ \sum_{e\in\mathcal{O}}\hat R_e\}$
is estimated from the observed jumps alone, with no knowledge of the
transition rates and no access to the hidden edges.

\begin{enumerate}
\item Record a stationary trajectory of duration $t$. On each observed
edge $e\in\mathcal{O}$ count the forward and backward jumps
$n_e^{+}$, $n_e^{-}$.
\item Form the net current and traffic
$\hat K_e = (n_e^{+}-n_e^{-})/t$,
$\hat T_e = (n_e^{+}+n_e^{-})/t$, the plug-in Schnakenberg cost
$\hat\sigma_e = \hat K_e\log[(\hat T_e+\hat K_e)/(\hat T_e-\hat K_e)]$,
and the quadratic remainder
$\hat R_e = \hat\sigma_e - 2\hat K_e^2/\hat T_e$.
Both directional counts must be positive; unresolved edges require
longer sampling or explicit statistical treatment, not an infinite
finite-sample certificate.
\item Estimate the current covariance by cutting the trajectory into
$B$ blocks of length $\tau = t/B \gg \tau_{\mathrm{corr}}$, forming the
per-block current vector $\hat K^{(b)}_{\mathcal{O}}$, and taking
$\hat\Sigma_{\mathcal{O}} = \tau\,
\mathrm{Cov}_b(\hat K^{(b)}_{\mathcal{O}})$.
\item Form the covariance term $2\,\hat K_{\mathcal{O}}^\top
\hat\Sigma_{\mathcal{O}}^+ \hat K_{\mathcal{O}}$, with the
pseudoinverse truncated below a singular-value cutoff that discards
directions the trajectory has not resolved.
\item Report $\widehat{\mathcal{B}}_{\mathcal{O}}^{\max}$ as above,
with sampling uncertainty.
\end{enumerate}

The block length $\tau$ and the singular-value cutoff control
statistical error, not the exact inequality. At fixed $\tau$,
ergodicity makes the block sample covariance approach
$\Sigma_{\mathcal O}^{(\tau)}
=\tau\,\mathrm{Cov}(\hat K_{\mathcal O}(\tau))$:
the block first and second sample moments converge to their stationary
expectations. This finite-window object need not equal $\Sigma_{\mathcal O}$.
Consistent long-time estimation requires growing blocks and many blocks,
$\tau\to\infty$ and $t/\tau\to\infty$, with suitable mixing and a cutoff
that consistently separates resolved covariance directions from the kernel.
The examples below use fixed long blocks and retain a small blocking bias.
Covariance inversion can bias the plug-in estimate upward, but blocking
and truncation also affect the bias. Stability under varying $\tau$ and
the cutoff should be checked; error bars alone do not make a plug-in
estimate a guaranteed finite-sample lower bound.

\section{Finite-sample test on a synthetic network}
\label{sec:finite-synthetic}

\begin{figure*}[tbp]
\includegraphics[width=\textwidth]{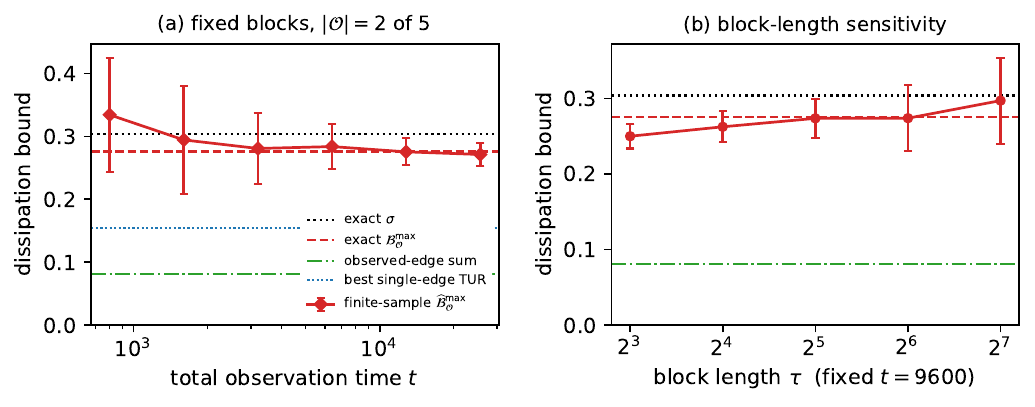}
\caption{Finite-sample test on a four-state two-cycle network,
observing a cotree of two of the five edges (cycle-observing).
(a)~The plug-in estimator $\widehat{\mathcal{B}}_{\mathcal{O}}^{\max}$
(red, mean $\pm$ standard deviation over $28$ trajectories) is compared
with the long-time bound (red dashed) at
$0.91\,\sigma$ (dotted $\sigma$), exceeding the scalar single-edge
TUR (blue, $0.51\,\sigma$) and the per-edge observed sum
(green, $0.27\,\sigma$). Fixed blocks $\tau=40$ retain a $2.16\%$
population bias; short-time overshoot illustrates plug-in sampling bias.
(b)~Sensitivity to block length $\tau$ at fixed $t=9600$.}
\label{fig:S-finite}
\end{figure*}

We test the estimator on a four-state, five-edge two-cycle network from
simulated trajectories, observing a cotree of two of the five edges.
This set is cycle-observing, so the completion is unique and the
covariance certifies the full density-contracted cycle energy.
The covariance level $0.9009\,\sigma$ from two of five edges is that of the optimized
multivariate current-fluctuation bound \cite{Gingrich2016,PLE2016}; the
cycle-observability result (Theorem~S6) identifies it as the full
density-contracted cost $2K^\top M_T K$ of all five edges, including
the three hidden ones, and the nonlinear correction
$\sum_{\mathcal{O}} R_e$ raises it to $0.9071\,\sigma$. The
scalar single-edge inference reaches $0.51\,\sigma$ and the per-edge
observed sum $0.27\,\sigma$. With fixed blocks $\tau=40$, the population
plug-in level is $0.26983$, compared with the long-time bound $0.27580$
($\sigma=0.30405$); this $2.16\%$ blocking bias persists as $t$ grows.
Figure~\ref{fig:S-finite} shows sampling scatter and block-length sensitivity,
not exact convergence to the long-time bound.

\section{Kinesin parameter and sampling conventions}

We use the six-state, low-product-concentration network and
Carter--Cross row of Table~I in Ref.~\cite{LL2007}, with the load
factors also discussed in Ref.~\cite{LL2009}. To impose exact balance
on rounded tabulated constants, we retain
$\kappa_{65}=0.02\,(\mu{\rm M})^{-1}{\rm s}^{-1}$ and set
\[
\kappa_{16}=
\frac{\kappa_{25}\kappa_{12}\kappa_{56}\kappa_{61}}
{\kappa_{52}\kappa_{21}\kappa_{65}K_{\rm eq}}
=0.0255102\,(\mu{\rm M})^{-1}{\rm s}^{-1},
\]
with $K_{\rm eq}=4.9\times10^{11}\,\mu{\rm M}$.
This preserves the model used in the figures; Table~I instead marks
$\kappa_{65}$ as balance-derived and supplies the rounded
$\kappa_{16}=0.02$. All other constants and head-symmetry relations
follow that row. At $1\,$mM ATP and $1\,\mu$M each of ADP and phosphate,
stall occurs at $7.19481\,$pN. The recovered quadratic bound is
$45.8168\,{\rm s}^{-1}$, whereas total entropy production is
$331.9165\,{\rm s}^{-1}$ in units of $k_{\rm B}$.
Multiplication by $k_{\rm B}T$ gives the power units used in the figure.
Stationary trajectories are divided into $50\,$s blocks; the fixed-window
covariance plug-in level is $45.6793\,{\rm s}^{-1}$, about $0.3\%$ below
the long-time quadratic bound. Both simulations use a relative
pseudoinverse cutoff of $10^{-3}$.

\bibliography{References}
\bibliographystyle{apsrev4-2}